\documentclass[10pt,preprint]{aastex}

\shorttitle{CCCP view of Trumpler 16} 
\shortauthors{Wolk et al.}

\newcommand{\minusone}{$^{-1}$}

\newcommand{\nh}{\mbox{$N_{\rm H}$}}

\newcommand{\skipthis}[1]{}

\newcommand{\ps}{{\rm s}^{-1}}

\newcommand{\erg}{{\rm ergs}}

\newcommand{\be}{\begin{equation}}
\newcommand{\ee}{\end{equation}}
\newcommand{\e}{et al.\ }

\begin{document}

\title{The Chandra Carina Complex Project View of Trumpler~16}


\author{Scott J, Wolk\altaffilmark{1}, Patrick S. Broos\altaffilmark{2}, Konstantin V. Getman\altaffilmark{2}, Eric D. Feigelson\altaffilmark{2}, Thomas Preibisch\altaffilmark{3},  Leisa K. Townsley\altaffilmark{2},  Junfeng Wang\altaffilmark{1}, Keivan G. Stassun\altaffilmark{4,5},  
Robert R. King\altaffilmark{6}, 
Mark J. McCaughrean\altaffilmark{6,7}, Anthony F. J. Moffat\altaffilmark{8} and Hans Zinnecker\altaffilmark{9} }

\altaffiltext{1} {Harvard--Smithsonian Center for Astrophysics,
       60 Garden Street, Cambridge, MA 02138, USA}

\altaffiltext{2} {Department of Astronomy \& Astrophysics, The Pennsylvania State University, 525 Davey Lab, University Park, PA 16802, USA}

\altaffiltext{3} {Universit\"ats-Sternwarte, Ludwig-Maximilians-Universit\"at, Scheinerstr.~1, 81679 M\"unchen, Germany}
       
\altaffiltext{4} {Department of Physics \& Astronomy, Vanderbilt University, Nashville, TN 37235, USA}

\altaffiltext{5} {Department of Physics, Fisk University, 1000 17th Ave. N., Nashville, TN 37208, USA}
      
\altaffiltext{6} {Astrophysics Group, College of Engineering, Mathematics, and Physical Sciences, University of Exeter, Exeter EX4 4QL, UK}

\altaffiltext{7} {European Space Agency, Research \& Scientific Support Department, ESTEC, Postbus 299, 2200 AG Noordwijk, The Netherlands}

\altaffiltext{8}{D\'epartement de Physique, Universit\'e de Montr\'eal, Succursale Centre-Ville, Montr\'eal, QC, H3C 3J7,
Canada}

\altaffiltext{9} {Deutsches SOFIA Insitute, Univ. of Stuttgart, Germany and NASA-Ames Research Center, USA}

\begin{abstract} 
Trumpler 16 is a well--known rich star cluster containing the eruptive supergiant $\eta$ Carin\ae\ and located in the Carina star-forming complex. 
In the context of the {\it Chandra Carina Complex Project}, we study Trumpler 16 using new and archival X-ray data.  
A revised X-ray source list of the Trumpler 16 region contains 1232 X-ray sources including 1187 likely Carina members. These are matched to 1047 near-infrared counterparts detected by the HAWK-I instrument at the VLT allowing for better selection of cluster members.  The cluster is irregular in shape. Although it is roughly circular, there is a high degree of sub-clustering, no noticeable central concentration and an extension to the southeast. 
The high--mass stars show neither evidence of mass segregation nor evidence of strong differential extinction. 
The derived power-law slope of the X-ray luminosity function for Trumpler~16 
reveals a much steeper  function than the Orion Nebula Cluster implying different ratio of solar- to higher-mass stars. We estimate the total Trumpler 16 pre-main sequence population to be $> 6500$
Class~II and Class~III X-ray sources.
An overall K-excess disk frequency of $\sim$ 8.9\% is derived using the X-ray selected sample, although there is some variation among the sub-clusters, especially in the Southeastern extension.
X-ray emission is detected from 29 high--mass stars with spectral types between B2 and O3.  
\end{abstract}

\keywords{ ISM: individual (Great Nebula in Carina) - open clusters and associations: individual (Trumpler 16) - stars: pre-main sequence - X-Rays: stars X-ray -- Facilities: Chandra, VLT}

\section{Introduction}

Trumpler 16 lies at the heart of the Carina Nebula region. It includes three main sequence O3 stars, the Wolf-Rayet star WR~25 and $\eta$ Carin\ae.  $\eta$ Carin\ae\ first became notable when it brightened significantly in the 1840s (Herschel 1847).  During that event an estimated $\ga 10$~M$_\odot$ of material and nearly 10$^{50}$ ergs of kinetic energy were injected into the host cluster (Smith \e 2003).  
Prior to the event, $\eta$ Carin\ae\ most likely dominated the energy budget of the cluster.
Since this event however, $\eta$ Carin\ae\ has been essentially cut off from the cluster due to the vast opaque shell surrounding it. 
This shell is surrounded by the Homunculus nebula (Gaviola 1950)  which itself is a subset of a massive HII region spanning several square degrees.  The 2.3 kpc distance to the cluster comes primarily from the expansion parallax of the Homunculus nebula around $\eta$ Carin\ae\ (Smith 2006, Davidson and Humphreys 1997). 

The stellar content and star formation history of Trumpler~16 has been studied in the optical band.  Early studies concentrated on the massive stars. Walborn (1971, 1973) identified 6 Henry Draper stars within Trumpler 16 and the neighboring cluster, Trumpler~14, which were more massive than any star  known at that time, leading to the introduction of  the O3 classification. This implied both a very high mass and young age for these cluster. Levato \& Malaroda (1982) and Morrell \e (1988) identified many more O stars in these clusters spectroscopically. Feinstein (1982) performed deep photomultiplier-based observations of about 70 cluster members as faint as $V \approx 14.$    An early CCD-based photometric study by Massey \& Johnson (1993) reached about 1 $M_\odot$.   This was  followed by Degioia--Eastwood \e  (2001) who presented optical photometry for over 560 stars in Trumpler 16.  They  found clear evidence of pre-main sequence stars in the region and argued for a mass-dependent spread in ages with  intermediate mass star forming continuously over the past 10 million years and high mass stars forming within the last 3 million years.  
  
 An inventory of Trumpler 16 includes 42 O stars  with the total radiative luminosity equal to  log ($L/L_\odot$) = 7.24,  a total mass loss equal to 1.08$\times 10^{-3}$ $M_\odot$ per year and mechanical luminosity of 6.7$\times 10^4~L_\odot$  in the wind (Smith 2006).
The region includes 12 small Bok globules including the  ``Keyhole Nebula'',  the ``Kangaroo Nebula'',  and ``The Finger''
(Smith \& Brooks 2008).  Whether or not these are sites of ongoing star formation remains an open question. The region is dominated by ionized gas emission. The strong 1.2 mm continuum is dominated by free--free emission, not cool dust (Brooks \e 2005).
 
\begin{figure}
\includegraphics[scale=0.7]{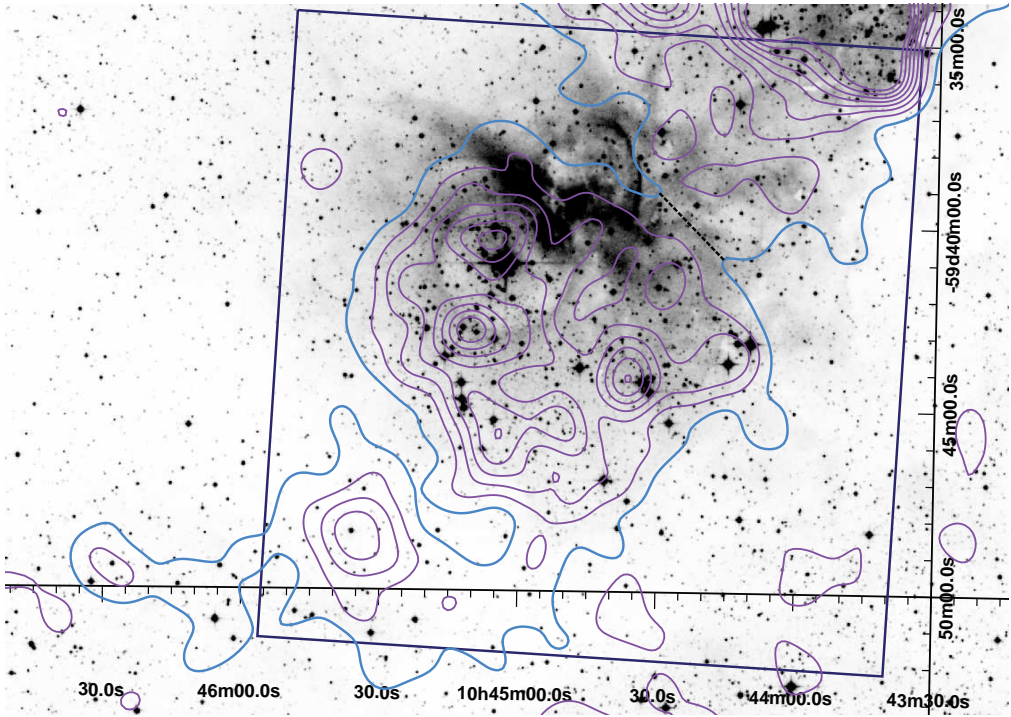}
\caption{Trumpler 16 cluster overview.  The background image is a far-red image from the Digitized Sky Survey (DSS2-I; squared scaling).  The square indicates the field of view of ObsID 6204.
The contours indicate an increase of source density of 1 X-ray source per 30\arcsec. The outer thick contour indicates the extent of the Trumpler 14 and 16 clusters.  $\eta$ Car is the bright star near the north-east X-ray source concentration. The dashed line indicates our boundary between Trumpler 16 and Trumpler 14 (to the Northwest). 
\label{mainimage}}
\end{figure}

In the X-ray regime, this is a very well studied cluster. Albacete--Colombo \e (2003)  observed this region with $XMM-Newton$ for 35 ks and detected 80 of the brightest sources, but this observation was badly limited by source confusion. Evans \e (2003) used 9.3 ks of early Chandra data  to study the hardness ratios of the hot stars in Trumpler 16.  The luminosity limit of that Chandra observation was about 7$\times 10^{31} \erg\ \ps$, typical of single O and early B stars.   Albacete--Colombo re-observed the cluster with Chandra for 90 ks.  They found 1035 sources and matched 660 to  2MASS counterparts (Albacete--Colombo \e 2008; AC08).  About 15\% of the X--ray sources with near IR counterparts were found to have infrared excesses indicative of an optically thick disk in the $K_S$ band.   While the AC08 study covered a square 17\arcmin\ by 17\arcmin\ ACIS-I field which included part of Trumpler~14,  the analysis presented by Feigelson \e (2011;  hereafter Paper I)  shows that  Trumpler~16 occupies only a portion of the ACIS-I field studied by AC08.    In the north, Trumpler~16 is roughly circular about 11.6\arcmin\  across while towards the southeast  it has an extension  about 10\arcmin\  by 6\arcmin\ (Figure~\ref{mainimage}). The structure, in part, is caused by an absorption lane that crosses the middle of the field. 
The highly structured nature of Trumpler 16 is qualitatively different from Trumpler 14 and Trumpler 15 which have a single central concentration (Ascenso et al. 2007, Wang \e 2011).  

The Trumpler 16 region was not re-observed as part of the {\it Chandra Carina Complex Project} (CCCP; Townsley \e 2011a); instead previous observations were re-analyzed following the prescriptions described in Broos \e (2010 and 2011).   The purpose of this paper therefore, is not a full discussion of the data, which are presented by AC08,  but instead to present the findings in the context of the full  X-ray analysis of the Carina complex.    We also invoke the new HAWK-I infrared observations (Preibisch \e 2011) for improved near-infrared counterpart information.   For the purpose of this paper, we define Trumpler 16 following the clustering analysis presented by  Paper I, in which Trumpler 16 is divided into seven sub clusters and a surrounding matrix of stars. 

In the next section, we will review the results of AC08 and compare those results with the re-analysis of the X-ray data using new techniques presented by Broos \e  (2010, 2011).   We then evaluate the global extinction due to dust, absorption due to gas, and luminosity properties of the cluster. Next, we study the spatial distribution of the stars concentrating on the several sub-clusters to examine whether the sub-clusters are real physical phenomena.   This paper will put little emphasis on high--mass stars except as they pertain exclusively to Trumpler~16, as these stars are discussed elsewhere (Naz\'e \e 2011, Gagn\'e \e 2011).   The A0-B3 stars in Tr~16 are examined in detailed by
Evans \e (2011),  candidate new OB stars are identified by Povich \e (2011), and Townsley \e (2011b) discuss the diffuse emission in the region.

\section{The Observations and Data Reduction}

The bulk of the X-ray data discussed here were taken prior to the CCCP, as part of a guaranteed time program (2006 August 31, PI S. Murray, 88.4 ks, ObsID 6402).
A comprehensive analysis of those data was provided by AC08, including discussion of sources in the Trumpler~14 region (to the north of Trumpler~16) and detailed analysis of individual sources that are X-ray bright.
In this study we limit our attention to the Trumpler~16 cluster, and we supplement the AC08 data with additional observations that partially overlap Tr~16 (ObsIDs 9482, 9483, 9488, 6578, and 4495), which are described and mapped by Townsley \e (2011a; [Table~1]).

We define Trumpler 16 as the region within the lowest contour of the kernel smoothed star surface density distribution shown in Paper I. This is identified by a continuous contour of density to the South and East and terminates in a narrow region separating Trumpler 14 from Trumpler 16 in the Northwest (Figure 1). Trumpler 16 includes the matrix of stars surrounding the 7 sub- clusters identified in Paper I. About 1/8 of the field of view of ObsID 6402 covers Trumpler~14 and about one--third of the area of ObsID 6402 lay outside of either cluster. Lower-density portions of Trumpler 16 to the South and East (which we will refer to as the Trumpler 16 Southeastern extension) were not included in the ObsID 6402 field of view. This includes the CCCP-cluster 14 (Paper I) which has been previously identified by Sanchawala et al. (2007a,b). This region has been covered by other observations as part of the CCCP program and so is included in the analysis presented here.

The 2MASS Survey, which has a completeness limit near the galactic plane of Ks Å 13.3 (Skrutskie \e 2006), was the only nearÐIR (NIR) data available to AC08.
Deep NIR observations obtained using the HAWK-I camera at the ESO VLT have recently become available and are used in this study to extend the NIR catalog to a completeness limit of Ks $\sim$  19 mag and an ultimate detection limit of Ks $\sim$ 21 (Preibisch \e 2011).
 
 The data reduction, source detection, and source extraction procedures applied to all the CCCP data are described by Broos \e (2011a).
We adopt the statistical classification of sources as likely Carina members, likely contaminants, or ``unclassified objects'' presented by Broos \e (2011b), understanding that the individual classifications are not guaranteed to be correct.
Column densities and absorption-corrected X-ray luminosities for individual stars were estimated using the photometric techniques of the XPHOT package (Getman et al. 2010).

Briefly, the CCCP source detection strategy was to nominate a liberal catalog of candidate point sources using multiple source finding algorithms and then iteratively extract those candidates, calculate for each a detection significance statistic (probability of the null hypothesis that all the X-rays found in the source aperture arose from the background), and prune candidates found to be not significant.
Our scientific goal to push for high sensitivity, accepting a non-trivial number of spurious detections (Broos \e 2011a).
The AC08 catalog was defined using a different algorithm (Palermo wavelet detection code, PWdetect; Damiani \e 1997) and more conservative thresholds that are expected to produce only  $\sim$ 10 spurious detections within the ACIS-I field of view.

Broos \e (2011a) discuss why estimating the number of false detections in the CCCP catalog or in the Trumpler~16 study region is not practical, and point out that such an estimate would be irrelevant for any analysis that further restricts the sample of stars, e.g., by requiring a ``likely member'' classification or requiring an estimate of X-ray luminosity from XPHOT.
However, an approximate {\em lower limit} on the number of legitimate X-ray sources in any sample can be obtained by tallying the number NIR counterparts identified.
Among the 1232 CCCP sources in our study area, 1067 (85\%) have NIR counterparts detected in at least one band. Among the 1187 classified as likely Carina members, 1047 (88\%) have NIR counterparts;  among the 885 with X-ray luminosity estimates, 804 (91\%) have NIR counterparts. Since we have confidence that X-ray sources bright enough for XPHOT photometry are real astrophysical sources, this limits the total fraction of false positives to a few percent of the faintest X-ray sources and hence should not effect any conclusion.

\section{Results: Global Considerations}

Structurally, Tr 16 is a roughly
circular  cluster about 11\arcmin\ across  (7.4~pc at 2.3~kpc).  We also consider the Southeastern extension to be part of the cluster (Figure~1). 
Table~1 enumerates the 1232 X-ray sources within the continuous contour with $> 1$ X-ray source per 30\arcsec\ kernel.
 Of these, the X-ray hardness and other criteria (Broos \e 2011) classify 1187 as probable members of the Carina complex, 11 as foreground objects, 2 as extragalactic  and 32 unknown.  
We find 392 sources not in the original catalog of  AC08.  Most of these are faint sources: the mean number of net counts in these new sources is 7 and the minimum is 2.3.  About 50 are found in the Southeastern extension which was not fully included in ObsID 6402.  The median energy of the previously detected sources is indistinguishable from the newly detected sources, $MedE \simeq 1.5$ keV.   The penultimate column of Table~1 lists the class of the X-ray source following Broos \e (2011) H0: unclassified; H1: source is a foreground main-sequence star; H2: source is a young star, assumed to be in the Carina complex; H3: source is a Galactic background main-sequence star; H4: source is an extragalactic source. The final column of Table~1 indicates the sub-cluster with which the source is identified (C3 = Sub-cluster 3 etc.) based on the nomenclature in Paper I.  We also identify those stars which are not identified with any single cluster as being part of the `matrix' of Trumpler~16.  The matrix stars in the Southeastern extension are identified separately as ``SEM". 

\subsection{Disk Fraction}
\begin{figure}
\begin{center}
\includegraphics[scale=0.7]{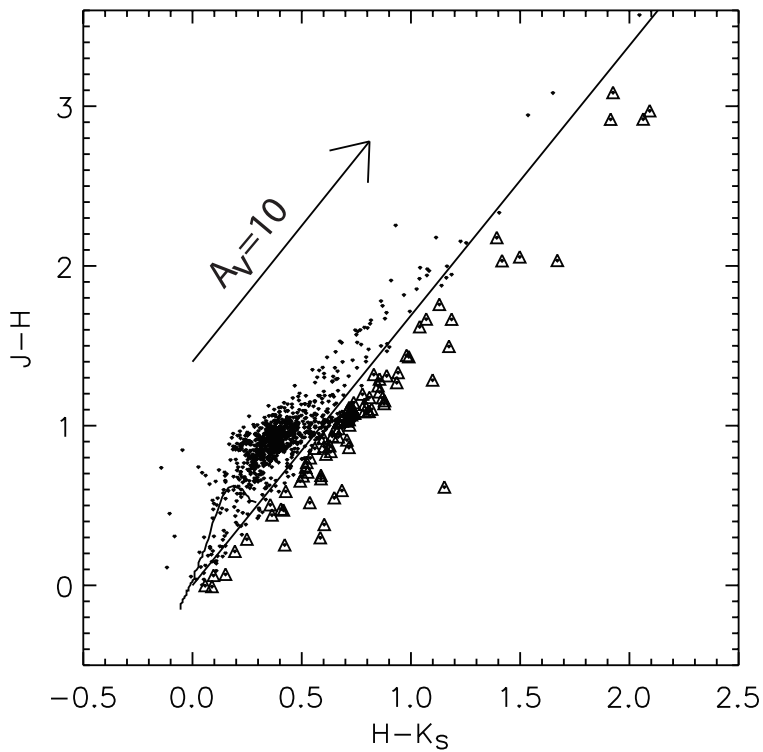}
\end{center}
\caption{Near-infrared color-color diagram of 1013 CCCP sources in Trumpler 16 (including  the Southeastern extension) using the HAWK-I data (Preibisch \e 2011).
 A reddening vector of 10 visual magnitudes is indicated.  The short curve in the lower-left indicates the nominal main sequence. The parallel lines indicate the reddening band for stars without optically thick disks in the $K_S$ band.
Triangles indicate stars at least 0.1 magnitudes to the right of the reddening band; these are probable disk systems with optically thick disks in the $K_S$ band. 
We find 9\% of the X-ray sources which are associated with probable cluster members have disks. Several of these are high mass stars.  Similar analysis was carried out separately on each subcluster within Trumpler~16. 
\label{CCD}}
\end{figure}

For the 1187 X-ray sources found to be probable members of the Carina complex, matches are found in the
HAWK-I photometric catalog for 1047. Almost all of those (1032 sources) are detected in all three $JHK$ bands.
The bulk of the X-ray sources matched have 12 $<K_S< $ 15.5 with wings extending  to both brighter and fainter sources.  
Of the 1032 ,probable members of the Carina complex with JHK detections,  1013 sources have errors less than 5\% in all bands. 
 Ninety of 1013  (8.9\% $\pm$0.9\%), have excesses consistent with an optically thick disk (Figure~\ref{CCD}).  This is a slightly higher rate than the 7.8\%, found for the full CCCP by Preibisch \e (2011) or the 6.9\% disk detections they found for the 529 X-ray sources located in the Trumpler~16 sub-clusters. 
The sources reported by Preibisch \e  are a sub-sample (only the sources which reside in sub-clusters) of the whole of Trumpler~16 which we covered here.
Also, we require a K$_S$ excess to be 10\%, this is higher than the Preibisch \e study, but is closer to the value used by AC08
who reported a relatively high disk fraction of  $\sim 15\%\pm2\%$.  

In AC08, the sample is restricted to the 339 2MASS sources with good colors in all three bands.   The true limiting factor for inclusion is the $K_S$ band magnitude which needs to be brighter than about 15.  Sources without a $K_S$ band excess are more likely to be excluded from the 339 stars sample, which will raise the disk fraction.  In the present study the IR  data are more complete than the X-ray data, so we do not expect a strong IR bias in favor of detecting stars with disks. 

\subsection{IR magnitude and X-ray Flux}

\begin{figure}
\includegraphics[scale=0.8]{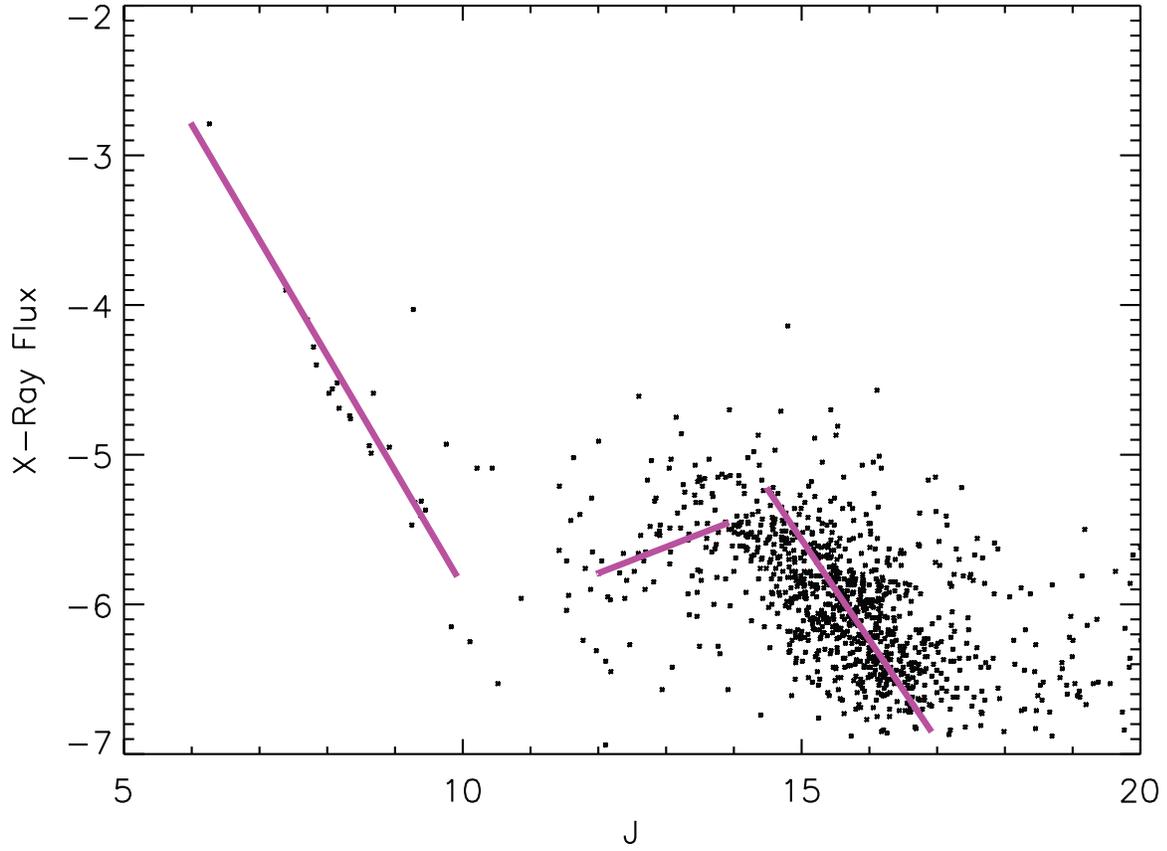}
\caption{The X-ray flux (in units of log photons sec\minusone cm$^{-2}$)  compared to the J band magnitude.   X-ray sources with $J<11$ (O and early-B stars) and within$14<J<17$ (pre-main sequence stars) show well-known correlations between X-ray and optical luminosities.  The weak anti-correlation at intermediate luminosities  (12$<$ J $<$ 14) is newly reported here.
\label{jvflux}}
\end{figure}

As seen in Figure~\ref{jvflux}, the X-ray fluxes of the sources are well correlated with $J$-band luminosities for $6<J<11$.
This relation appears to reverse between $12<J<14$ and reappears in the range $14<J<18$.
Below $J$ =18 the flux distribution appears flat.  
Most of this can be readily interpreted as follows.  Given the 2.3 kpc distance and a roughly 3 Myr age, objects with  $6<J<11$, are primarily high--mass stars, which generate X-rays through shocks in unstable radiatively driven winds (Lucy 1982).  Both the luminosity and the shock speeds scale with mass down to early-B stars (See the Naz\'e \e 2011 in this issue).    Similarly, objects with $14<J<18$ are pre--main sequence (PMS) G, K through mid--M stars that are known to have their bolometric luminosity correlate with their X-ray luminosity.  This is seen for example, in the Chandra Orion Ultradeep Project (COUP) study of the Orion Nebula Cluster (Preibisch \e 2005). For these stars, roughly 0.02\% of their luminosity is emitted in X-rays although the overall fraction can be higher during X-ray flares.  Below $J$=18, typical PMS stars are too faint to be detected in these observations and only those caught during strong flares are seen. 
 
 The behavior between $12<J<14$ is curious. For the 2.3 kpc distance and $\sim$ 3 Myr age these are 
 Mid-B through A stars, which are not efficient X-ray producers since they have weak winds, but $\sim 30-60\%$ of such stars in  Trumpler~16 and COUP were detected in X-rays (Evans \e 2011; Stelzer \e 2005).   Both Evans \e and Stelzer \e concluded that high energy emission from these sources originates in unseen companions.   It has not previously been reported that the X-ray flux became brighter as the stars became bolometrically fainter.  Previous samples were probably too small to detect this effect.  If the X-ray emission was indeed the result of unseen companions, the implication is that the higher mass primaries (B5-A5) typically have lower mass companions than the primaries below A5. 
 
To quantify these effects, we performed a piecewise least-squares linear regression and applied nonparametric correlation measures to the data in Figure~\ref{jvflux}.  The fitted slope is $m=-0.76 \pm 0.05$ for $6<J<10$, $m=-0.30 \pm 0.02$ for $14.5<J<17$ but reverses direction to $m=0.18 \pm 0.08$ for $12<J<14$ (these lines are plotted in Figure~\ref{jvflux}).  Kendall's $\tau$ correlation coefficient (Kendall 1938) gives  $\tau=-0.86$ for the bright stars, $-0.41$ for the faint stars, and 0.21 for the intermediate brightness stars.  The probability for the correlation of the intermediate stars is $P \sim 0.5$\%, roughly equivalent to a 3$\sigma$ effect. 

Because we found the effect surprising we repeated the tests, allowing the limits on the intermediate J magnitude band to vary by up to 0.5 mag.  We also calculated probabilities in various terms including the Spearman $\rho$ correlation coefficients (Spearman 1904) for each these are -0.70 for the bright J band sources, -0.30 for the fainter J band sources and 0.13 for the intermediate case.  In all cases, there is a weak positive correlation in the middle range. Further,  a similar pattern has been seen in the Trumpler~14 and Trumpler~15 clusters within the $\eta$ Carina cluster (Evans \e $in~prep$). No correlation is seen when the entire CCCP sample is used, however this sample covers relatively broad range of ages and conditions. 

\subsection{Near--IR Extinction} \label{NIR_AV.sec}

We originally calculated extinction to each source following {\it \~A}$_V=(H-K_S-0.1)\times 13.7$ (Preibisch \e 2011).\footnote{
Following Preibisch \e we use the expression  ``{\it \~A}$_V$" to indicate extinction calculated using this simple scaler relation. We use 
``{\it A}$_V$" to indicate extinction measurements obtained by individually fitting a given stellar color + extinction to a model color in this case using Siess \e (2000).} 
Preibisch \e note that {\it \~A}$_V$ calculated in this manner may not be accurate for all masses: the first term assumes stellar photospheric color $H-K_S=0.1$ appropriate for 3~Myr stars between 1-2 $M_\odot$ (Siess \e 2000).  Below 1 $M_\odot$, photospheric $H-K_S$ exceeds 0.1 and extinction is over-estimated, and above  2 $M_\odot$, extinction is underestimated.  This can lead to estimates of negative extinction.
To minimize these effects, we only estimate extinction for stars with no evidence of optically thick disks at $K_S$ and $J$ magnitudes between 14.5 and 16.5 as these are likely to have masses of
 1-- 2 $M_\odot$.
 
The second term is the proportionality factor characteristic of the extinction law.
We note the derived reddening law measured for this region appears to deviate from the typical ISM value of R= 3.1,  with derived values being between 3.8 and 5 with possible spatial variations (Naz\'e \e 2011, Povich \e 2011,  Gagn\'e \e 2011, Smith 1987, Th\'e 1980, Herbst 1976, Forte 1978, Feinstein \e 1973).  However, using only optical data, Turner \& Moffat (1980) found R = 3.2 throughout the Carina region. He we us {\it \~A}$_V/E_{(B-V)}$=4.0 (Povich \e 2011) which leads to the constant values $k$=13.7.  A higher value of $R_V$  would lower the value of $k$.  Given these constraints, we find the mean {\it \~A}$_V=3.8$ with 25\% and 75\% quartiles at 2.9 and 5.0, respectively, for the Trumpler 16 CCCP sample of likely Carina members.  

AC08 calculated extinction by using a $K_S$ vs. $J-H$  color-magnitude diagram, individually dereddening stars until the location of the star intersected the isochrone for a 3 Myr cluster PMS  (isochrones from Siess \e 2000).  They found a mean reddening of  {\it A}$_V$=3.6$\pm$2.4 mag, acknowledging  that their estimates of {\it A}$_V$ could be in error by up to 0.7 mag due to the relatively high photometric errors in the 2MASS data.   While overall this method is more precise than that used by Preibisch \e (2011), the errors may be higher than estimated due to their unconfirmed assumption that all stars are exactly 3 Myr old. 
Furthermore,  we expect our data to be deeper and hence capture more highly extinquished stars than the 2MASS sample. 
Another important aspect for the comparison of HAWK-I and 2MASS is the spatial resolution:
about 2 arcsec for 2MASS versus about 0.6 arcsec for HAWK-I. Many "point-like" 2MASS sources
are resolved into several components in the HAWK-I images.
Other effects such as photometric variability of the young
stars, unresolved binary companions, and small infrared excesses 
(that are too small to move the star out of the main-sequence
reddening band in the color--color diagram) will all add both random and systematic errors 
into the extinction measurements.  

We examined a sub-sample of about 100 stars with non-degenerate\footnote{When calculating infrared extinctions for 3 Myr stars, the reddening vector crosses isochrones multiple times for stars with masses between about 2.2-8.0 $M_\odot$.  For these stars, IR data alone are not conclusive and we identify the reddening determinations as degenerate.} reddening  extinctions measured by  AC08. 
We find an offset $\Delta$ = 0.58 mag 
between the mean value for {\it \~A}$_V$ obtained here and the mean {\it A}$_V$ published by AC08. The latter tends to show less extinction.   There is considerable scatter between the two  {\it \~A}$_V$ estimates, $\sigma$ = 1.0.  When we restrict the sub-sample further to 65 stars with extinction corrected magnitudes of $13<K_S<14$, which would place their masses between 1 and 2 $M_\odot$,  the difference and dispersion are reduced to $\Delta = 0.4$ and $\sigma = 0.8$. This indicates that extending the 
simple correction used by Preibisch \e beyond its designed range caused some of the scatter.  We then calculated extinctions for each source, following the methods described by  AC08, but using the HAWK-I data and found that large scatter and systematic offsets around {\it A}$_V \simeq 0.5$ persist.

Considering all factors  -- ranges in age, measurement errors, mass limitations, dust particle size distribution --  the level of
agreement  between the different methods appears reasonable.  Since the assumption of intrinsic $H-K_S = 0.1$ is accurate to within 3\% for stars with masses $2.2>M_\odot >0.8$ which dominate the X-ray distribution, the extinction estimator of Preibisch \e (2011) seems appropriate for the CCCP Carina member sample exclusive of the O, B and A stars. 

\begin{figure}
\begin{center}
\includegraphics[scale=0.7]{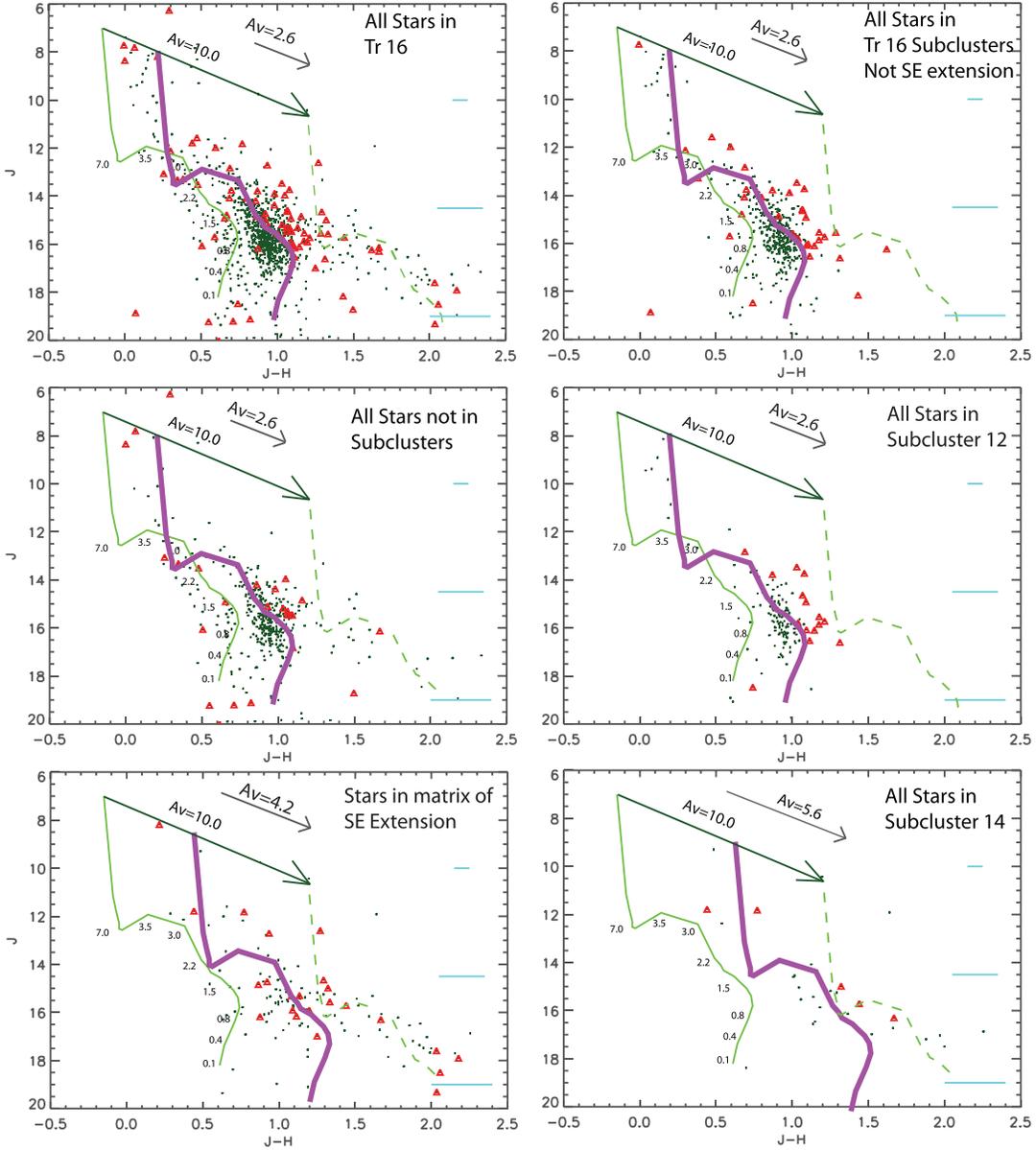}
\end{center}
\caption{Near-infrared color-magnitude diagrams for Trumpler 16 and several of its sub--structures.
The green solid line indicates a 3 Myr isochrone derived from Siess \e (2000) set at a distance modulus of 11.8.  The dashed green line is the same isochrone with 10 $A_V$ of extinction applied.  The thick purple line is an approximate fit of the isochrone to the data with only the extinction being allowed as a free parameter.  The short arrow in the upper middle of each frame shows the derived extinction for each region.  The triangles indicate stars with disks as derived from the IR-color--color diagram.
All of the regions in the main part of Trumpler~16 fit well to an extinction of $A_V=3.3$.  The stars in the Southeastern extension average about 150\% this extinction, and its Sub-cluster 14 is even more absorbed. The bulk of the disked stars are more absorbed than the cluster mean. The cyan horizontal lines indicate rough color error bars.
\label{CMDs}}
\end{figure}

 In Figure~\ref{CMDs} we show the $J$ vs. $J-H$ color-magnitude diagrams for Trumpler~16 and several of the sub-clusters.
 The solid (green) line on the left hand side of the plot 
 is an isochrone from  Siess \e (2000) assuming an age of 3 Myr and a distance modulus of 11.81.  We can globally fit the 
 extinction to the cluster, by dereddening the ensemble of stars with $13<J<17.5$ in steps of 0.1 $J$ magnitudes and measuring the $\chi^2$ residuals versus the  3 Myr isochrone model.
 We restricted the sample to stars without evidence for optically thick disks. 
 The best fit was found for an extinction value of {\it A}$_V$=3.3 with interquartile confidence band of $2.3- 3.8$. 
 When the sample is restricted to 1--2 $M_\odot$ likely Carina members, the mean becomes $A_V = 4.1$ with interquartile range $3.4 - 4.9$.

\subsection{X--Ray Absorption} 

Individual X-ray luminosity and absorption were calculated  for X-ray sources in the direction of Trumpler~16 using the XPHOT package (Table \ref{xphot.tab}; Getman \e 2010).
The concept of the XPHOT method is similar to the use of color--magnitude diagrams in optical and infrared astronomy, with X-ray median energy replacing color index and X-ray source counts replacing magnitude.   Using non-parametric methods, one can estimate both apparent and intrinsic broadband X-ray fluxes and soft X-ray absorption from gas along the line--of--sight to X-ray sources. Apparent flux is estimated from the ratio of the source count rate to the instrumental effective area averaged over the chosen band. Absorption, intrinsic flux, and errors on these quantities are estimated from comparison of source photometric quantities with those of high signal--to--noise spectra that were simulated using spectral models characteristic of 
low--mass pre--main sequence stars. 
In the original paper, the results were compared with spectroscopic analysis of sources in M~17 (Broos \e 2007).
For stars with median total band energy $>1.7$ keV, Getman \e (2010) show that fluxes measured by XPHOT agree with fluxes obtained by spectral fitting to within a factor of $\sim$ 1.5 for sources with more than 50 counts, and within a factor of $\sim$ 4 for sources with as few as 10 counts.  The total band covers the 0.5 -- 8.0 keV energy range.

For the Trumpler~16 sample, 347 sources could not be assigned a flux or temperature. Eighty percent of these had less than 10 counts and all but six had less than 30 counts. The final six had photon distributions inconsistent with the model of a one-temperature $\sim$ 1.5 keV corona.  The remaining  885 sources had X-ray luminosity and absorption calculated.  For sources with between 10--20 counts, the median error in total flux in total flux is about 35\% and the statistical error in \nh\ is also about 35\% (for median energy of 1.5 keV).
In the lowest count bin, 5--7 counts, the median error in total flux in total flux is about 65\% and the statistical error in \nh\ is also about 45\% (for median energy of 1.5 keV).  There is also a systematic error on the \nh\ value due to low effective area below about 500 eV. The
systematic error is about 15\% at 1.5 keV but approaches unity for sources below 1 keV.  The converse of this is 
that ACIS spectral resolution is not fine enough to discriminate log \nh\  values significantly below 22.0 (Getman \e 2010).  Below log \nh =21.6 systematic errors dominate over statistical errors with a median statistical error in log \nh\ of about 0.3 and systematic errors as high as log \nh =1.15. 
Above log  \nh =21.6 statistical errors dominate with a median statistical error in log \nh\ of about 0.2 but statistical errors still as high as log \nh=1.0. It is cautioned that log \nh\ values below 21.6 are less robust than high values of \nh.

The mean \nh\ value derived from XPHOT results here is essentially identical (within 10\%) to that found by  
AC08 using other methods.  However, we find a systematic bias between XPHOT and XSPEC \nh\ values for the stronger X-ray sources in Trumpler~16 in the sense that for log \nh (cm$^{-2}$) $\le$ 21.75 , XPHOT gives a larger value and for log \nh (cm$^{-2}$) $>$ 21.75  XSPEC gives a higher value.  At the extreme values,  the differences can be 0.75 dex. This finding is consistent with the result in Getman \e (2010) wherein the observed median energies of stellar sources in M~17 were found to have a simple linear relation to the log \nh\ for values of log \nh (cm$^{-2}$) $>$ 21.6. The XPHOT derived absorptions were deemed unreliable for 205 sources with log \nh $<$ 21.6 cm$^{-2}$ and below a median energy of 1.7 keV, due to degeneracy in the median energy -- log \nh\ relationship used by XPHOT.   

\begin{figure}
\begin{center}
\includegraphics[scale=0.7]{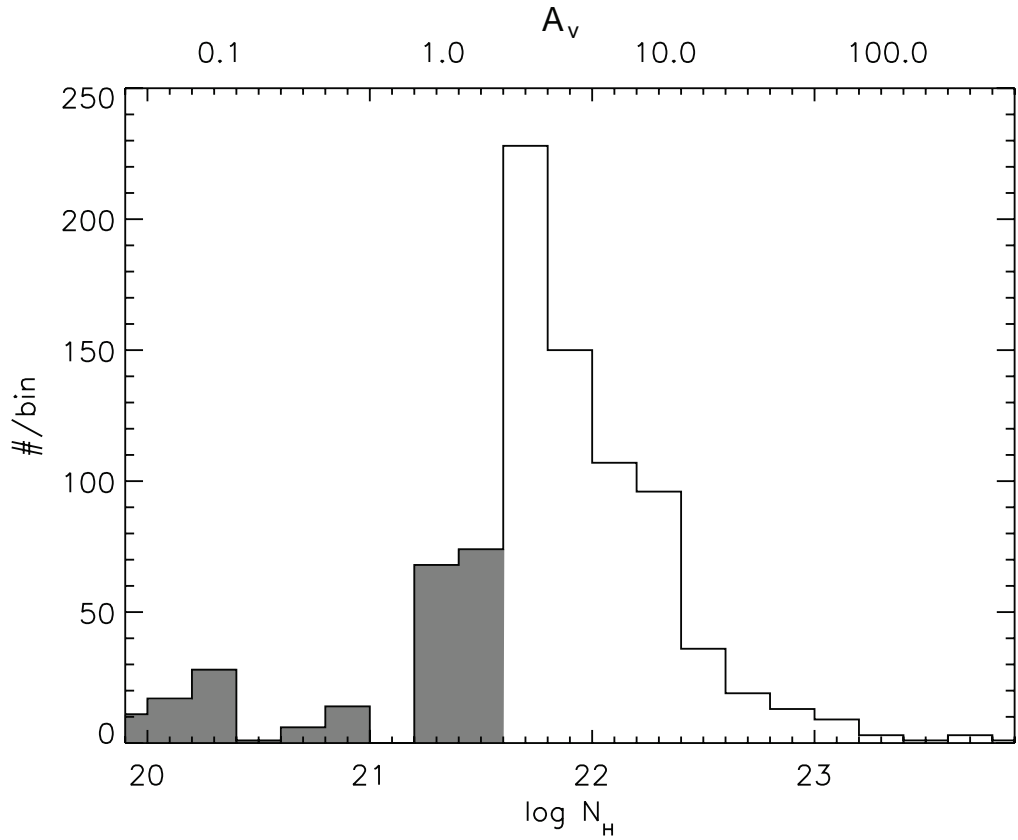}
\end{center}
\caption{Histogram of log \nh\ of 687 X-ray sources in Trumpler~16 as measured using the XPHOT method.  The top axis gives the approximate  optical extinction ({\it A}$_V$) using a conversion ratio of \nh/$A_V =1.6\times 10^{21}$ (Vuong \e 2003).  The histogram is in grey below log \nh= 21.6 to indicate the less robust nature of these measurements (see text). 
\label{NH}}
\end{figure}

Figure~\ref{NH} shows the distribution of absorption columns found by XPHOT for the sources in the direction of Trumpler~16.  The distribution is highly structured, especially when compared to Figure~7 of  AC08. 
 About 5\% of the sources have log \nh\ $<$ 21; some of these may be Galactic field foreground stars misclassified as likely Carina members.  About 15\% of the CCCP likely Carina members have 21.2 $<$ log \nh (cm$^{-2}$)  $<$ 21.6, leading to a strong peak in the distribution of sources at log \nh\ $\simeq 21.7$ cm$^{-2}$, with a similar shoulder around 22.0 $<$ log \nh (cm$^{-2}$) $<$ 22.3.  A small number of sources have higher absorptions in the range 22.3 $<$ log \nh (cm$^{-2}$) $<$ 23.4; some of these may be embedded Carina protostars, while others may be misclassified extragalactic sources.

If the very low absorption tail (log \nh(cm$^{-2}$) $<$ 21.0) is attributed to foreground contaminants,  then the distribution of X-ray absorptions has a strongly peaked unimodal distribution very similar to the distribution of absorptions derived from IR photometry (Figure~\ref{CMDs}). Using the conversion \nh\ = $1.6 \times 10^{21} A_V$ cm$^{-2}$ obtained by Vuong et al. (2003), the peak of the Trumpler 16 X-ray absorption distribution is equivalent to $A_V \simeq 3$ mag, in agreement with the value $A_V \simeq 3-4$ mag obtained from the near-IR color-magnitude diagrams (\S~\ref{NIR_AV.sec}).  Both the X-ray and IR absorption distributions have a sparse tail of extinctions found going out to {\it A}$_V > 30$ mag.

\subsection{X-ray Luminosity Distribution} 

\begin{figure}
\begin{center}
\includegraphics[scale=0.7]{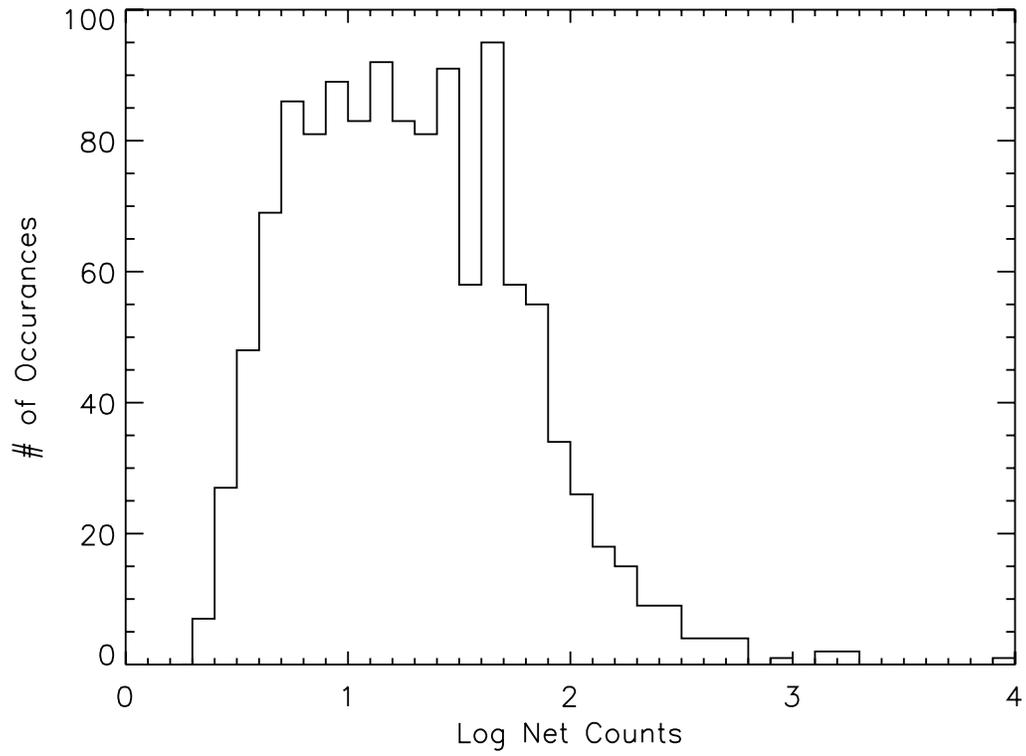}
\end{center}
\caption{Histogram distribution of the net counts from the 1232 X-rays sources detected in the Trumpler 16 region.
\label{net_counts}}
\end{figure}

As most of the CCCP sources in Trumpler 16 are too faint in the X-ray band for direct spectral modeling (Figure~\ref{net_counts}), the XPHOT technique is used to scale the observed count rate to broad-band luminosity with a correction for soft X-ray absorption.  Getman \e (2010) find that most XPHOT total-band fluxes are within $\pm 20$\% of values determined with XSPEC for the absorption ranges typically seen in the Trumpler 16 region.  They estimate the errors in total-band source fluxes to be better than 60\%, 50\%, 30\%, and 20\% for net count strata 7--10, 10--20, 20--50, and $>$50 counts, respectively, with a small systematic bias.  In Trumpler~16, nearly 500 of the 885 sources with fluxes and luminosities measured by XPHOT have less than 7 net counts. The simulations estimate less than 70\% errors for these sources. 

As noted by Feigelson \e (2005), the X-ray luminosity function (XLF) is the product of the initial mass function and the correlation between mass-age and X-ray luminosity.  This correlation is well-studied in the COUP and Taurus young stellar populations (Preibisch \e 2005, Telleschi \e 2007).  The X-ray luminosity functions of the ONC, IC~348, NGC~1333 and other young clusters appear similar, suggesting that the XLF shape may be roughly ``universal'' (Wang \e 2008).  For the youngest clusters, differential evolutionary effects
appear minimized and the XLFs of the clusters are well fitted by a log-normal function with 
$<$log L$_X>$(erg s$^{-1}$) = 29.3 and $\sigma$=1.0 (Feigelson \e 2005).
Wang \e (2008)  find some deviations from precise agreement for different clusters; for example, a steeper slope in the XLF of  M17 and a more shallow slope for the Cep OB3b cluster is seen.  
Meanwhile M17 has nearly 3 times the stars of the ONC while Cep B less than half the unobscured population of the ONC
(Broos \e 2007, Getman \e 2005, 2006).
The implication is that more massive clusters may have a steeper slope to their luminosity function.

\begin{figure}
\begin{center}
\includegraphics[scale=0.7]{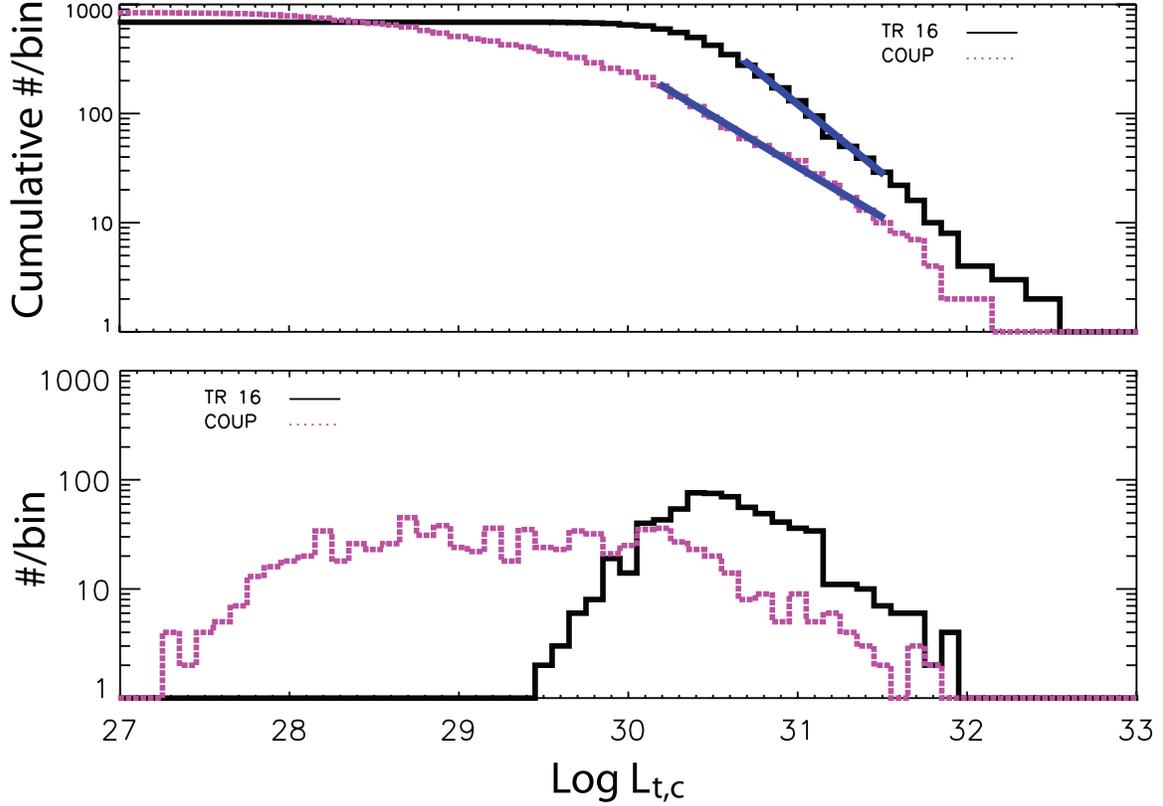}
\end{center}
\caption{
Top: Power-law fits to the Trumpler~16 data from log $L_{t,c}$= 30.7-31.5 and to the COUP data from log $L_{t,c}$= 30.2-31.5 are shown by blue line.  The Trumpler~16 distribution has a slope of $\Gamma = -1.27$, while the COUP data have a slope of  $\Gamma = -0.92$
Bottom: Histogram of  absorption-corrected,
total-band (0.5-8keV) X-ray luminosities of all 687 X-ray sources in Trumpler~16 (black, solid) for which XPHOT could calculate luminosities and which are not in the SE extension. These are compared to the COUP sample (magenta, dotted) of 839 sources from the ONC. 
\label{XLF}}
\end{figure}

 Figure~\ref{XLF} compares the XLF of Trumpler~16 (excluding the SE extension) with the XLF of the COUP data (Getman \e 2005). 
To determine the XLF of Trumpler 16, the first step is to estimate the completeness of the X-ray data.  Figure~\ref{XLF}b shows a histogram of the luminosity as derived by XPHOT of 687 X-ray sources observed in Trumpler~16 (this is down from 885 because we are excluding the SE extension which has  different properties as will be discussed in \S~4.).  
Visual examination of  Figure~\ref{XLF}b shows a drop off  starting at log $L_{t,c}$= 30.5 
($L_{t,c}$ = absorption corrected luminosity in the total band covering 0.5 - 8.0 keV). This value is consistent with the CCCP
completeness limits discussed by Broos et al. (2011a), which vary strongly with off-axis distance and with absorption.
The implication is that  incompleteness in the distribution occurs before this point, mostly before log $L_{t,c}$= 30.7. 

  As shown in Figure~\ref{XLF}a, we fitted the cumulative XLF of Trumpler~16 (excluding the SE extension) with a power-law between log $L_{t,c}$= 30.7 - 31.5 and find a slope $\Gamma=-1.27$.  This slope was sensitive to the lower luminosity cut--off used at the level of 0.05 in slope for a 20\% change in the luminosity limits;  $\Gamma=-1.13$ if we brought the lower cut--off down to $L_{t,c}$= 30.3  which is clearly incomplete.  The upper cutoff of $L_{t,c}$= 31.5 is essentially the brightest cool star. 
 This slope is steeper than a similarly measured slope for the COUP data which is found to be  $\Gamma=$-$0.93$ for  log $L_{t,c}$ =30.2 -- 31.5. The slope of the COUP data is less sensitive to the lower luminosity cut--off and is only effected at the level of 0.01 for a 20\% change in the luminosity limits.

To estimate the total number of sources in the cluster, we simply compare the total number of sources in Trumpler~16 to the number in the COUP sample in the range from log $L_{t,c}$= 30.7 - 31.5  which should be dominated by cool stars. 
The COUP sample contains 51 sources in this range, while the Trumpler~16 sample has 255 for a factor of 5($\pm0.5$) more.
Given an estimate for the total population of the ONC from the COUP as 1300 X-ray sources  (Getman \e 2005) we estimate the total population of Trumpler~16 at 6500 $\pm650$ if observed to a similar X-ray luminosity limit.   Further, Hillenbrand \& Hartmann (1998) estimate the overall ONC population to be 2800.  If this is correct, then the X-ray sources represent less than 50\% of the total cluster membership which may be as high as 14,000.

\section{The Sub-clusters within Trumpler~16}

Paper~I defines Trumpler~16 as a region in which the surface density of CCCP likely Carina members smoothed with a 30\arcsec\ Gaussian kernel (FWHM=0.8~pc) exceeds 1 source per kernel.  Within this region, Paper~I
identified seven sub-clusters within Trumpler~16 with sub-cluster surface density exceeding three sources per 30\arcsec\ Gaussian kernel. 
Overall, Paper~I finds 31 small X-ray selected groups of probable Carina members in the CCCP.
Trumpler~16 contains 8 of these groups, including one in the Southeastern extension, which we refer to as sub-clusters. No single sub-cluster exceeds 15\% of the total number of sources in the Trumpler~16 cluster.  
The exact number and shape of these sub-clusters varies depending on the smoothing kernel the density factor. 
 This is unlike Trumpler~14 and 15 which are each dominated by a single, central concentration. 
 
\begin{figure}
\begin{center}
\includegraphics[scale=0.7]{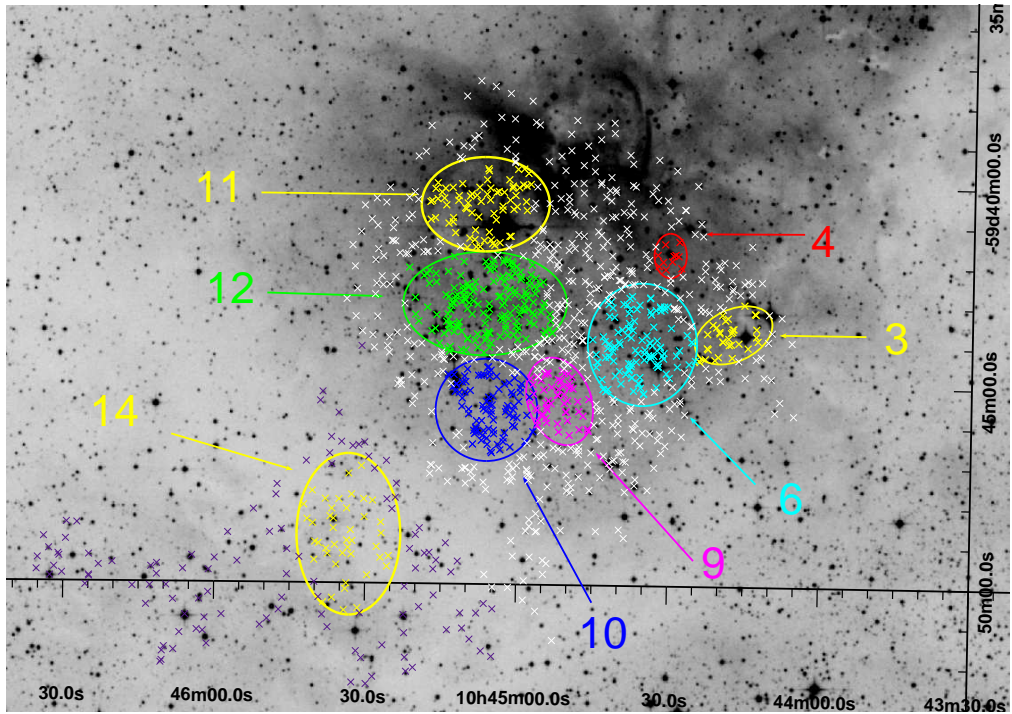}
\end{center}
\caption{The sub-clusters within Trumpler 16 with X-ray sources noted by $\times$ symbols.   Each sub-cluster is labeled  with designations from Paper I and shown by an approximate ellipse. In the color version the various sub-clusters are indicated by color as well. Members of the matrix are indicated in white and occasionally cross into region ovals as the latter are approximated.   The background image is the same as Fig.~\ref{mainimage}, but linearly scaled.
\label{subclust}}
\end{figure}

In this section, we examine whether the sub-clusters are physically distinct units or simply statistical fluctuations, and we compare their properties. 
The sub-cluster identification for each source is given in Table~1 and Figure~\ref{subclust}.  In the figure, we have approximated each sub-cluster as an ellipse to aid in the calculation of geometric parameters.   For each sub-cluster we have assessed extinction as 
{\it \~A}$_V$ for stars of $K_S <$ 13.  We then applied the extinction to the observed $K_S$ magnitude and the distance to produce a K-band luminosity function (KLF).  As discussed above, the actual values of individual stars are not reliable and may be in error by up to 0.2 $K_S$ mag.,  but the structure of the distribution should be representative.  To quantify the structures, we have identified the quartile values of absolute $K_S$  and {\it \~A}$_V$ for all sub-clusters. 

In order to determine if the sub-clusters truly stand out from the background cluster population of Trumpler~16, we first look at the 506 X-ray sources which form the matrix of Trumpler~16.  These are cluster members, but not coincident with any specific sub-cluster, nor the Southeastern extension. 
Of the 438 sources with good mid-IR colors, 6.4\% ($\pm$ 1.2\%)  show IR excesses consistent with an optically thick disk.  This is more than 2$\sigma$ less than the cluster as a whole and may indicate that these dispersed stars are more evolved than the global population.
The mean extinction value of {\it \~A}$_V=3.8$ is indistinguishable from the cluster as a whole.

\subsection{Sub-cluster 3}
Sub-cluster 3 is the westernmost of the Trumpler~16 sub-clusters.  When a larger smoothing kernel is used, 
it appears as a low density extension of Sub-cluster 6. 
It contains 33 X-ray sources and has  a spatially averaged source density of 32 src pc$^{-2}$.  Twenty--seven of the X-ray sources  have been matched with HAWK-I sources; almost all of these are between 1 and 2 $M_\odot$. About $15\% \pm 7$\% of the IR detected X-ray sources have disks. The mean {\it \~A}$_V$ is 3.8 similar to the cluster as a whole. 
The massive supergiant Tr 14 Y 398 (spectral type O3~Iab) as well as WR~25 (HD~93162, WN) lie on the western edge of the sub-cluster.    

\subsection{Sub-cluster 4}
Sub-cluster 4 is the smallest of the Trumpler~16 sub-clusters with only 11 X-ray sources. The  
 spatially averaged source density of 36 src pc$^{-2}$ is about the same as Sub-cluster~3.  Of the eight X-ray sources matched to HAWK-I, two have $K_S$ excesses consistent  with an optically thick disk. The mean {\it \~A}$_V$ is 3.5.  
The B1~V star CPD$-$59${\rm ^{o}}$2581 lies near center of the sub cluster, and the similar B1~V CPD$-$59${\rm ^{o}}$2574 at the northwestern edge of the sub-cluster.  Neither of these stars were detected in X-rays.

\subsection{Sub-cluster 6}
Sub-cluster 6 is a centrally condensed sub-cluster and the second largest in Trumpler~16 overall with 109 sources and an average source density of 45 src pc$^{-2}$.  
Of the 88 sources with good mid-IR colors, 6.8\% $\pm$ 2.8\%  show IR excesses from an optically thick disk, consistent  with the global population. Sub-cluster 6 has a similar absorption to the previously discussed regions with mean {\it \~A}$_V = 3.6$ and  75\% quartile {\it \~A}$_V = 4.1$.  
There are three O stars in this sub-cluster, the O3.5 V+ O8 V double HD 93205 and an O5 V HD 93204. These are displaced by 0.25 pc to the southwest of the cluster center.

 \subsection{Sub-cluster 9}
Sub-cluster 9 is the western half of a bow-tie shaped enhancement toward the southern part of Trumpler~16.  The  53 sources appear uniformly distributed with no internal concentrations and a high spatially averaged source density of 48 src pc$^{-2}$.  
The X-ray sources here are relatively faint, about half of the sources in the region required the more sensitive source detection procedure described by Broos \e (2010) and were not found by AC08. This is about twice the usual fraction of newly identified sources. 
Of the  45 sources with good mid-IR colors, only 2 (4\%$\pm$ 3\%)  show IR excesses consistent with an optically thick disk. One of these is exceptionally faint, $K_S=19.5$ and hence the colors are quite dubious. The mean extinction calculated for the region as a whole is typical of other regions with {\it \~A}$_V =$3.7.  There are no high-mass stars in this region. 

\subsection{Sub-cluster 10}
Sub-cluster 10 is the eastern half of  the bow-tie shaped enhancement toward the southern part of Trumpler~16.  The 82 sources appear uniformly distributed with no strong internal concentrations and a high spatially averaged source density of 42 src pc$^{-2}$.  The X-ray luminosity is more typical here, 20\%  of the sources required the more sensitive source detection procedure described by Broos \e (2010).  Of the  74 sources with good mid-IR colors, only 4 (5.4\%$\pm$ 2.7\%)  show IR excesses consistent with an optically thick disk. Two of these are  faint, $J >18.1$ and hence the colors have errors of about 35\%. Thus there are only two bona-fide disk systems in this subcluster. 
 The mean extinction calculated for the region as a whole is typical of other regions with mean {\it \~A}$_V=$3.7.  
 Four late O stars in the region are all detected in X-rays. The most massive,  O7 V  star CPD$-$59${\rm ^{o}}$2626, lies near the subcluster center, while the others lie towards the east and southeast.
 
 Although sub-clusters 9 and 10 are in contact, they have different mass distribution. Sub-cluster 9 has fewer higher mass stars.  Both clusters have relatively low numbers of stars with optically thick disks in the  $K_S$ band.

\subsection{Sub-cluster 11}
$\eta$ Carin\ae\ and 6 other high-mass stars reside in sub-cluster 11.
The 71 X-ray sources include five of the high mass stars. Overall the X-ray sources are  centrally condensed 
with a peak at 10:45:06 $-$59:40:25, about 0.3~pc from $\eta$ Carin\ae.  The spatially averaged source density of 27 src pc$^{-2}$ is among the lowest of any region.   This may be due, in part, to reduced point source sensitivity in the immediate vicinity of $\eta$ Carin\ae\ and the associated X-ray nebula. 
Of the  57  sources with good mid-IR colors, 4  show IR excesses consistent with an optically thick disk (7\%$\pm$ 4\%).  
The mean extinction calculated for the region as a whole is low, $<${\it \~A}$_V> =$2.9.  

\subsection{Sub-cluster 12}
\begin{figure}
\begin{center}
\includegraphics[scale=0.6]{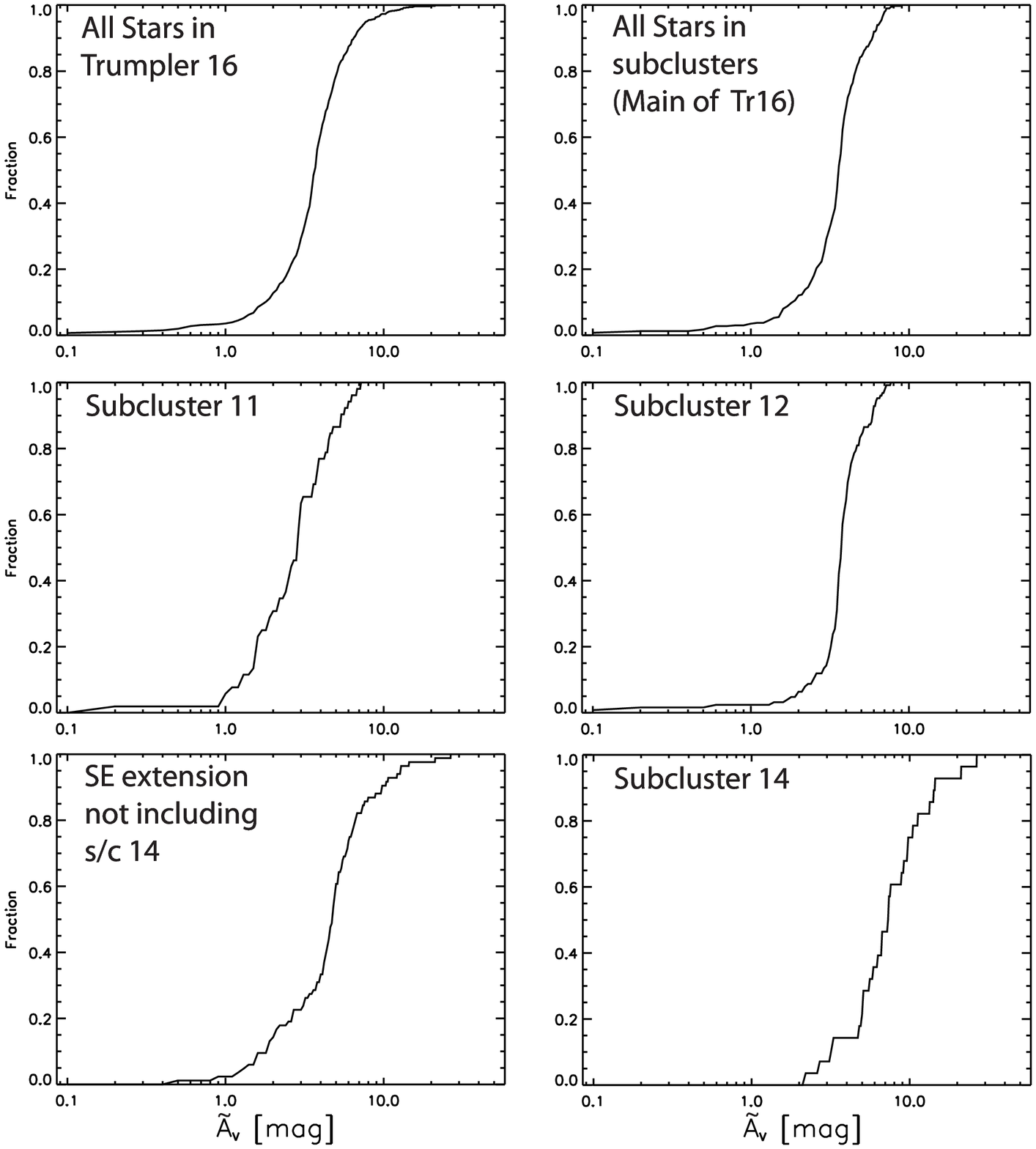}
\caption{The extinction functions for several regions within Trumpler 16. }
\label{AV}
\end{center}
\end{figure}

Sub-cluster 12 is immediately to the south of sub-cluster 11, centered at 10:45:10.6, $-$59:42:54.  It is the most centrally condensed and populous sub-cluster of Trumpler~16 with 166 X-ray sources.  It includes six of the nine most massive stars in Trumpler ~16.   There are two high-mass stars within 15\arcsec\ of the center, both of which are early B stars, not O stars.  The most massive star complex in the cluster, CPD$-$59${\rm ^{o}}$3310,  contains two high--mass stars, an O6 V + a B2 sub-giant, and  is located about 2\arcmin\ to the southeast of the cluster center. An additional candidate OB star at the center of sub-cluster 12 is identified
by Povich \e (2011).  This is the only one of 94 new OB star candidates in the CCCP located within the main area of Tr~16  
(three were identified around the periphery and five in the Southeastern  extension).\footnote{There is also a strong selection effect as the  Spitzer VelaÐCarina survey was highly constrained when observing near  $\eta$ Carin\ae.}
The spatially averaged source density is 45 src pc$^{-2}$.  Of the 142 sources with good mid-IR colors, $10.6\% \pm 2.7\%$ show IR excesses consistent with an optically thick disk.  All but one of these are found at relatively high extinction ({\it A}$_V > $2.6), indicating they likely lie behind a source of extinction within the sub-cluster.  Further, the cumulative extinction distribution is very steep, with 60\% of the X-ray sources with  $3.4<$ {\it \~A}$_V < 4.2$.  All the disked stars are among the most absorbed 15\%.
It appears there is a thin cloud  with {\it \~A}$_V \sim 1$ near the front of this sub-cluster. Figure~\ref{AV} shows the extinction functions of several regions within Trumpler~16, with sub-cluster~12 showing the steepest distribution.
 
\subsection{The Southeastern Extension and Sub-cluster~14}
To the southeast of the main body of Trumpler~16, separated by a dust lane, is a continuation of the density distribution which we identify with Trumpler~16.   This region was first recognized by Sanchawala \e(2007a, b) who identified 10 X-ray sources in this region using the first two $Chandra$ observations of the Carina nebula (ObsIDs 1249 and 50).  The region is distinguished by high extinction, $<${\it A}$_V>$= 4.2, and a steep KLF indicating a young age. Using the larger CCCP mosaic, we find this extension covers about 40\% of the area of the main cluster and contains 156 X-ray sources.  The bulk of this extension was outside the field of ObsID 6402 and has somewhat lower overall exposure time, about 60 ks rather than 90 ks.    Hence, a lower surface density of sources is expected.  Even in light of the difference in exposure times, it is clear that conditions are different in this region. The disked fraction is much higher than the rest of the cluster, $19\pm4$\%. 

Within the Southeastern extension, there is  a density enhancement identified as sub-cluster 14 in Paper~I.  This corresponds with the location of the 10 X-ray sources identified by  Sanchawala \e (2007a, b).  We find 40 stars within this sub-cluster and a very high extinction,   $<${\it \~A}$_V> =$7.4.  Sub-cluster 14 is identified as a more embedded region  within the Southeastern  extension.  This sub-cluster also has the highest disked fraction, 21$\pm8$\%, of any sub-cluster with a significant number of stars.  The eclipsing O5.5 + O9.5 binary V662 Carin\ae\ is located  within about 15\arcsec\  (0.1~pc) of the cluster center making sub-cluster 14 the only one of the eight sub-clusters with a dominant O star so close to its geometric center.    Four of the five OB candidates identified by Povich \e (2011) in the Southeastern extension lie within sub-cluster 14.  
Three rank among the seven most luminous of all the OB star candidates in the CCCP.
    
The matrix of stars surrounding sub-cluster 14 is composed of 116 X-ray sources in the region, including the O4 V star LS 1886  located in the eastern portion of the extension. Removing sub-cluster 14 from the Southeastern extension,the region still stands out with  high extinction  $<${\it \~A}$_V> =$4.8 and a high disked fraction of 18$\pm4$\%. 

 \subsection{Discussion of Sub-clustering} 
The metrics for each sub-cluster are summarized in Table~\ref{summetric}. 
 It is clear that the main body of Trumpler~16 (subclusters 3, 6, 9, 10, 12 and the surrounding matrix) is different from the Southeastern extension.  The extinction is nearly 3 magnitudes greater in the Southeastern extension and the disked fraction is more than twice that of the main body. 
Within the Southeastern extension, sub-cluster 14 is distinguished  by an additional 2 magnitudes of extinction in the V band and a higher disked fraction.  The disk fraction can be used as a proxy for age (Haisch \e 2001).  While the age calibration 
is sensitive to the sensitivity of the survey, disked fraction appears to decrease nearly linearly in time -- indicating that the Southeastern extension is about 80\% the age of the main body of Trumpler~16.

Within the main body of Trumpler~16, the sub-clusters appear to be dynamically distinct structures.  Some have $> 100$ stars with surface densities $> 3$ times that of the outer portions of the matrix, too rich to arise from statistical fluctuations.  On the other hand, except for chance difference in line-of-sight absorptions with respect to Carina clouds, the X-ray selected stars in the sub-clusters are similar to each other and to the stars in the matrix.  Stars in all of the sub-structures outside of the SE extension have disk fractions consistent with $\sim 7$\%.
The stars in the matrix tend have a slightly lower disk fraction than the stars in the sub-clusters but the significance is 
$< 2\sigma$.  

The massive O stars, both main sequence and supergiant, do not lie at the cores of the subclusters.  Few young stellar clusters are documented to have widely distributed massive stars; the best example may be NGC~2244, the central cluster in the Rosette Nebula (Wang \e 2008).

\section{The Relation of the High--Mass Stars to the Sub-clusters}

X-ray emission is detected from 29  stars in Trumpler~16 that were classified with spectral types earlier than B2 (Skiff  2010).  This includes $\eta$ Carin\ae\ itself, which was detected, but not cataloged 
by Broos \e (2011) due to its high degree of pile-up (saturation in the ACIS CCD detection) which makes characterization difficult. The remainder include WR~25 and 19 O stars, many of which are in multiple systems.  
In addition, there are eight B stars in the range from B0.5 to B2 which were detected in X-rays,  six of these with less than 15 net counts. 

There are 21 early-type stars which were not detected in the X-ray observations. Most of the non-detections are from B0V to B2V in spectral type. This indicates a very sharp drop off in X-ray activity across the O/B divide as all the O stars were detected.  There were also a few early--type stars not detected in X-rays which were also not considered to be main sequence either. 
All of these are in the far south and all but emission object Hen 3-480 are considered to be part of the Southeastern extension. 
We list the detections in Tables 4 and 5 and the non-detections in Table~6. More comprehensive studies of early--type stars are  the topics of separate papers in the framework of the CCCP (e.g., Naz\'e \e 2011; Gagn\'e \e 2011). Candidates from Povich \e (2011) are excluded from this discussion as well since followup is still required to establish their spectral types and confirm their OB status.

Following the finding of AC08 that OB stars are less absorbed than their late type counterparts ({\it A}$_V$=2.0  for OB stars versus 3.6 for GK stars), we examine the extinction to the X-ray detected early--type stars. Twenty-five of these have matches in the HAWK-I photometry catalog.  We calculated the extinction by comparison to the $J-H$ and $J-K_S$ colors of high--mass stars which are expected to be nearly constant at $-0.15$ and $-0.2$ respectively, in the absence of extinction.  We find a range of extinctions from $A_V \sim 0.85 - 8.9$.  The average $A_V$ is about 2.5  but this is dominated by a single outlier (MJ 224) with $A_V \sim 8.75$.  The mean extinction may be the more appropriate metric at $A_V \sim 2.1$. However the dispersion is high, $\sigma=1.0$ even excluding the outlier. Hence, the extinction of the high--mass stars is generally lower than that of the  low--mass population.  The single high mass member of the Southeastern  extension detected in X-rays has $A_V \sim 4.5$ consistent with membership in Sub-cluster 14.

Remarkably, the spatial distribution of the high--mass stars does not follow the sub-clustering seen in the lower mass stars.   None of the seven sub-clusters in the main body of Trumpler~16 has an O star within 0.2~pc of the cluster center;
this includes sub-cluster 11 -- the host of $\eta$ Carin\ae.  The centroid of the X-ray sources in this sub-cluster is located 0.7\arcmin\ (0.5 pc) from $\eta$ Carin\ae.  
The unclustered matrix has the same fraction of high-mass stars to lower mass stars.   Only Sub-cluster~14 in the Southeast extension appears centered on a high mass star.  

\section{Trumpler~16 in Context}

In Paper~I, it is shown that the Carina nebula has very complex clustering properties including many sparse groups, a few small clusters such as Bochum 11 and the ``Treasure Chest," as well as three large clusters, Trumpler~14, Trumpler~15, and Trumpler~16. This 
range of stellar groupings is seen in other large star forming regions such as the Orion A cloud, which includes the ONC, OMC 2/3, Lynds 1641 North, Lynds 1641 South, as well  as numerous small agglomerations.  But comparison of the three largest clusters implies something other than size matters.  

In Table~\ref{comp} we provide a direct comparison of the clusters.  The data sets are not analyzed in identical ways but the methods are comparable. Trumpler~14 appears to be the youngest of the three clusters as determined by a variety of metrics including the location of the stars on the color--magnitude diagram and the disk fraction.  It is also possesses the most X-ray sources and is most centrally condensed. In this Table we calculated the total number of stars in Trumpler~14 based on the total mass estimate of 9000--11,000 $M_\odot$  of stars by Ascenso \e (2007) and then estimating the mean stellar mass to be 0.8 $M_\odot$ following Hillenbrand \& Hartmann (1998).   Ascenso \e estimate the distance to the Carina nebula to be 2.8~kpc, not  the 2.3~kpc used here, so the total number of stars should be estimated to be perhaps 10\% larger.   
The densities shown in Table~\ref{comp} were all estimated using a 2.3~kpc distance. 
Meanwhile, Trumpler~15 is the oldest, and lowest mass with a less dense core than Trumpler~14.  Trumpler~14 and Trumpler~15 are well fitted by King models (King 1962) with a core radius of about 0.7 pc containing about 30--35\% of the stars. 

Trumpler~16 does not fit into any simple relationship with its two neighbors to the north. It is nearly identical in mass (stellar numbers) to Trumpler~14. The main body of Trumpler~16 and the southeastern extension appear to bracket Trumpler~14 in age and the overall stellar density of the two is nearly the same.  The number of high-mass stars traces the total estimated population of all three clusters. Of course. some high mass stars have gone supernova -- more in the older clusters.  But Trumpler~16 is different. It has no identifiable core and its high mass population is fully distributed. While Ascenso \e find no mass segregation in Trumpler~14 either, the high mass stars trace the centrally condensed overall population. So overall it would appear that Trumpler~16 is the result of a different mode of star formation. 

In this mode the gas within Trumpler~16 appears to have been collected into several lumpy groupings within which the mass clumps were not smoothly distributed.  Before a single dynamical timescale, perhaps a triggering event occurred which caused all the clumps to collapse contemporaneously.  This is in contrast to the rather organized nature of the other two large clusters.
Nonetheless, the mode of star formation which produced Trumpler~16 created a high-mass to low mass star ratio nearly identical to that seen in the other clusters and an XLF  nearly identical to that of Trumpler~15, which is well described by a simple King model.  This implies that the IMF and the XLF are robust to different modes of star formation, assuming that the observed IMF and the XLF are not significantly altered by the evolution of cluster members.

 \section{Conclusions}

The Chandra ACIS observations of the Trumpler~16 cluster, portions of which were previously studied by Albacete-Colombo et al.\ (2008), were presented in the framework of the CCCP study.  Our analysis shows that Tr~16  is an irregularly shaped cluster. It is not highly centrally condensed but rather breaks up into several sub-clusters.   With the benefit of the deep HAWK-I data to identify faint counterparts to the X-ray sources we find no mass segregation in the cluster as a whole. Neither is there apparent mass segregation within the individual sub-clusters. In fact, none of the sub-clusters appear centered on a high mass star with the exception of sub-cluster~14.
Other results are summarized as follows:

\begin{enumerate}

\item  There are 1187 X-ray sources in the total CCCP sample which are classified as likely members of the Trumpler~16 cluster.  
Positional coincidence matching yields a total of 1047 HAWK-I near-IR counterparts, 1013 of these have three band detections with errors less than 5\%. 

\item The Trumpler 16 cluster has a roughly circular shape about 11\arcmin\ across  (7.4~pc at a distance of 2.3~kpc).  Within this outline are seven sub-clusters identified in Paper I.  We also discuss a less densely populated, more embedded and younger Southeastern  extension to the cluster which is about 10\arcmin\  long running east to west  and about 5\arcmin\ north to south (6.7~pc $\times$3.4~pc).
 
 \item We confirm the well-established correlations between X-ray flux and near-IR magnitude for high-mass stars and pre-main sequence G and K stars. We also find a weak anti--correlation  between X-ray flux and near-IR luminosity for intermediate mass stars.  
  
\item  The X-ray luminosity function for stars in Trumpler 16 is compared to the COUP Orion Nebula Cluster XLF. 
Trumpler~16 shows more low X-ray luminosity and  solar mass stars and proportionally fewer high luminosity 
(intermediate/high mass) stars than in ONC. 
We estimate the total number of X-ray sources brighter than  $L_{t,c}$= 27.5, the nominal limit of the COUP, to be about 5 times the number of Class~II and Class~III sources in the ONC area covered by the COUP survey. This estimate excludes secondary companions.

\item  The locations of X-ray detected stars in Trumpler~16 in the near-IR color-magnitude diagram is consistent with a population of 1-3 Myr PMS stars. The extinctions range from near 0 to over 20 in $A_V$.  There is some spread with 2.0 mag of extinction at $V$ typically separating the first and fourth quartile. 

\item   We derive an overall $K$-band excess disk frequency of 8.9$\pm$ 0.9\% using the X-ray selected sample.  Excluding the Southeastern extension the disk frequency is about 7\%. Both rates are significantly larger than the rate found in Trumpler~15 -- the oldest of the Carina rich clusters -- of 3.8 $\pm$ 0.7\%   (Wang \e 2011).  

\item We study the seven density enhancements within the main body of the cluster, some of which present unique characteristics. The stellar characteristics of the sub-clusters are very similar.   The matrix and the subclusters each contain roughly half the stellar population of Trumpler~16. No mass segregation is seen (i.e., the massive stars are not concentrated in the cluster cores). The pre-main sequence disk fraction found in the subclusters is 8.4$\pm$ 1.4\%  which is  consistent with 6.4$\pm$ 1.2\% found in the surrounding matrix of stars.  The exception is the  Southeastern extension which has a disk fraction more than a factor of 2 higher.  Absorption properties differ, particularly in the Southeastern extension that is heavily extinguished.  

\item In addition to the high extinction and higher disk fraction,  the Southeastern extension was also found to possess some of the most bolometrically luminous newly identified O star candidates in the region.   Taken together, the high disk fraction, high extinction, and luminous O stars  are evidence that the Southeastern extension is younger than the core region of Trumpler~16.

\item  We detected 29 previously known high mass stars including $\eta$ Carin\ae, WR~25, and main sequence stars with spectral types ranging from B2 to O3. Most B0 to B2 stars in this region were not detected.  We find marginal evidence that high--mass stars are less absorbed than lower mass stars.

\item The overall structure of Trumpler~16 differs greatly from that of Trumpler~14 and Trumpler~15. Nevertheless the XLF of 
Trumpler~15 and Trumpler~16 are nearly identical.

\end{enumerate}

\acknowledgements
We thank the referee for many useful comments. S.J.W. is supported by NASA contract NAS8-03060 (Chandra). 
This work is supported by Chandra X-ray Observatory grant GO8-9131X (PI: L. Townsley) and by the ACIS Instrument Team contract SV4-74018 (PI: G. Garmire), issued by the Chandra X-ray Center, which is operated by the Smithsonian Astrophysical Observatory for and on behalf of NASA under contract NAS8-03060.   AFJM is grateful to NSERC (Canada) and FQRNT (Quebec) for financial aide. The near-infrared observations were collected with the HAWK-I instrument on the VLT at Paranal Observatory, Chile, under ESO program 60.A-9284(K).  This research has made use of the SIMBAD database and the VizieR catalogue access tool, operated at CDS, Strasbourg, France. 

\section{References}
\noindent Albacete Colombo, J.~F., M{\'e}ndez, M., \& Morrell, N.~I.\ 2003, \mnras, 346, 704 \\
Albacete-Colombo, J.~F., Damiani, F., Micela, G., Sciortino, S., \& Harnden, F.~R., Jr.\ 2008, \aap, 490, 1055 (AC08)\\
Alexander, D.~M., et al.\ 2003, \aj, 126, 539 \\
 Ascenso, J., Alves, J., Vicente, S., Lago, M.T.V.T.\ 2007  \aap, 476, 199\\
Brooks, K.~J., Garay,  G., Nielbock, M., Smith, N., \& Cox, P.\ 2005, \apj, 634, 436 
\\
Broos P.~S., \e 2011a CCCP Catalog paper 
\\
Broos P.~S., \e 2011b CCCP Paper - A Naive Bayes Source ClassiÞer for X-ray Sources
\\
Broos, P.~S., Townsley, L.~K., Feigelson, E.~D., Getman, K.~V., Bauer, F.~E., \& Garmire, G.~P.\ 2010, \apj, 714, 1582 
\\
Broos, P.~S., Feigelson, E.~D., Townsley, L.~K., Getman, K.~V., Wang, J., Garmire, G.~P., Jiang, Z., \& Tsuboi, Y.\ 2007, \apjs, 169, 353 
\\
 Damiani, F., Maggio, A., Micela, G., \& Sciortino, S.\ 1997, Statistical Challenges in Modern Astronomy II, 417 
\\
 Davidson, K., \& Humphreys, R.~M.\ 1997, \araa, 35, 1 
\\
 DeGioia-Eastwood, K., Throop, H., Walker, G., \& Cudworth, K.~M.\ 2001, \apj, 549, 578 
\\
 Evans, N.~R., Seward, F.~D., Krauss, M.~I., Isobe, T., Nichols, J., Schlegel, E.~M., \& Wolk, S.~J.\ 2003, \apj, 589, 509 
\\
Feigelson, E.~D, et al.\ 2005, \apjs, 160, 379 
\\ 
Feigelson E. \e 2011, CCCP Clustering paper (Paper~I)
\\
Feinstein, A., Marraco, H.~G., \& Muzzio, J.~C.\ 1973, \aaps, 12, 331 
\\
Feinstein, A.\ 1982, \aj, 87, 1012 
\\
 Forte, J.~C.\ 1978, \aj, 83, 1199 
\\
Gagn\'e, M., \e 2011 CCCP Hot Star paper
\\
Gaviola, E.\ 1950, \apj, 111, 408 
\\
Getman, K.~V., Feigelson, E.~D., Grosso, N., McCaughrean, M.~J., Micela, G., Broos, P., Garmire, G., \& Townsley, L.\ 2005, \apjs, 160, 353 
\\
Getman, K.~V., Feigelson, E.~D., Townsley, L., Broos, P., Garmire, G., \& Tsujimoto, M.\ 2006, \apjs, 163, 306 
\\
 Getman, K.~V., Feigelson, E.~D., Broos, P.~S., Townsley, L.~K., \& Garmire, G.~P.\ 2010, \apj, 708, 1760 
\\
Haisch, K. E., Lada, E.~A., Lada, C.~J.\ 2001 \apj, 553, L153
\\
 Herbst, W.\ 1976, \apj, 208, 923 
\\
Herschel, J.~F.~W., Sir 1847, {\it Results of Observations Made During the Years 1834, 5, 6, 7, 8 at the Cape of Good Hope} (London: Smith, Elder and Co.)  
\\
 Kendall, M.\ 1938," {\it Biometrika} 30, 81
\\
 King, I. 1962, \aj, 67, 471
\\
Levato, H., \& Malaroda, S.\ 1982, \pasp, 94, 807 
\\
Morrell, N., Garcia, B., \& Levato, H.\ 1988, \pasp, 100, 1431 
\\
Naz\'e, Y., \e 2011 CCCP O star  paper 
\\
Povich, M. S. \e 2011 CCCP New OB star  paper 
\\
 Preibisch, T., \& Feigelson, E.~D.\ 2005, \apjs, 160, 390 
\\
 Preibisch, T., et al.\ 2005, \apjs, 160, 401 
\\
Preibisch, T., \e 2011, CCCP HAWK-I paper
\\
Sanchawala, K., Chen, W.-P., Lee, H.-T., Chu, Y.-H., Nakajima, Y., Tamura, M., Baba, D., \& Sato, S.\ 2007, \apj, 656, 462 
\\
Sanchawala, K., et al.\ 2007, \apj, 667, 963 
\\
 Siess, L., Dufour, E., \& Forestini, M.\ 2000, \aap, 358, 593 
\\
Skiff, B.~A.\ 2010, VizieR Online Data Catalog, 1, 2023 
\\
Skrutskie, M.~F., et  al.\ 2006, \aj, 131, 1163 
 \\
 Smith, N., \& Brooks, K.~J.\ 2008, Handbook of Star Forming Regions, Volume II, 138 
\\
Smith, N.\ 2006, \apj, 644, 1151 
\\
Smith, N., Gehrz, R.~D., Hinz, P.~M., Hoffmann, W.~F., Hora, J.~L., Mamajek, E.~E.,  \& Meyer, M.~R.\ 2003, \aj, 125, 1458 
\\
 Smith, R.~G.\ 1987, \mnras, 227, 943 
\\
 Spearman, C.\ 1904,  {\it Amer. J. Psychol.}, 15, 72
\\
 Stelzer, B., Flaccomio, E., Montmerle, T., Micela, G., Sciortino, S., Favata, F., Preibisch, T., \& Feigelson, E.~D.\ 2005, \apjs, 160, 557 
\\
 Th\'e, P.~S., Bakker, R., \& Tjin A Djie, H.~R.~E.\ 1980, \aap, 89, 209 
\\
Townsley, L., \e 2011a CCCP Introduction paper 
\\
Townsley, L., \e 2011b CCCP Diffuse emission paper
\\
 Walborn, N.~R.\ 1971, \apjl, 167, L31 
\\
Walborn, N.~R.\ 1973, \apj, 179, 517
\\
 Wang, J., Townsley, L.~K., Feigelson, E.~D., Broos, P.~S., Getman, K.~V., Rom{\'a}n-Z{\'u}{\~n}iga, C.~G., \& Lada, E.\ 2008, \apj, 675, 464 
\\
Wang, J., Townsley, L.~K., Feigelson, E.~D., Getman, K.~V., Broos, P.~S., Garmire, G.~P., \& Tsujimoto, M.\ 2007, \apjs, 168, 100 
\\
 Wolk, S.~J., Winston, E., Bourke, T.~L., Gutermuth, R., Megeath, S.~T., Spitzbart, B.~D., \& Osten, R.\ 2010, \apj, 715, 671

\skipthis{

}

\begin{deluxetable}{lrrrrrrl}
\tabletypesize{\normalsize}
\tablecaption{Basic Properties of Chandra ACIS Point Sources in the Trumpler~16  Region \label{basic}}
\tablewidth{0pt}
\tablehead{
\colhead{Seq. \#} &\colhead{Designation}&\colhead{R.A.} &\colhead{Dec.} &
\colhead{Net}&  \colhead{X-ray Cnt} &\colhead{Class} &\colhead{Sub-} \\
\colhead{~}              &\colhead{CXOGNCJ}&\colhead{J2000.} &\colhead{J2000} &
\colhead{X-ray Cts} &  \colhead{err} & \colhead{~} &\colhead{cluster} 
 }
\startdata
5277	&104405.29$-$594543.4	&161.022048	&$-$59.762068&	24.9&5.8	&H2 &	Matrix\\
5294	&104405.79$-$594353.6	&161.024162	&$-$59.731567&	12.4&4.7		&H2&	Matrix\\
5409	&104407.86$-$594315.7	&161.032779	&$-$59.721036&	73.0&9.6		&H0& Matrix\\
5416	&104408.06$-$594522.4	&161.033601	&$-$59.756223&	36.8&6.7		&H2&	Matrix\\
5466	&104409.04$-$594538.7	&161.037671	&$-$59.760751&	49.5&7.4		&H2&	Matrix\\
5493	&104409.75$-$594338.7	&161.040642	&$-$59.727435&	19.8&5.1		&H2&	Matrix\\
5496	&104409.80$-$594448.0	&161.040853	&$-$59.746691&	49.6&7.4		&H2&	Matrix\\
5534	&104410.39$-$594311.1	&161.043323	&$-$59.719771 & 4609.4&162.6	      &H2&	Matrix\\
5541	&104410.47$-$594352.7	&161.043650 	&$-$59.731332&	15.3&4.8		&H2&	Matrix\\
5576	&104411.23$-$594445.7	&161.046796	&$-$59.746034&	12.6&4.0		&H2&	Matrix \\
5609	&104411.89$-$594414.8	&161.049575	&$-$59.737468&	14.1&4.4		&H2&	Matrix\\
5629	&104412.43$-$594212.6	&161.051825	&$-$59.703517&	18.9&4.9		&H2&	Matrix\\
5638	&104412.53$-$594351.3	&161.052242	&$-$59.730934&	21.9&5.2		&H2&	C3\\
5647	&104412.84$-$594333.1	&161.053541	&$-$59.725883&	16.3&4.5	      &H2&	C3\\
5649	&104412.86$-$594344.6	&161.053617	&$-$59.729078&	162.1&13.0 	      &H2&	C3\\
\enddata
\tablenotetext{a}{Table 1 with complete notes is published in its entirety in the electronic edition of the Astrophysical Journal. A portion is shown here for guidance regarding its form and content.}
\tablenotetext{b}{Column 1: CCCP X-ray catalog sequence number (Broos \e 2010). Column 2: IAU designation. Columns 3,4: Right ascension and declination for epoch J2000.0 in degrees. Column 5: Net X-ray events detected in the source extraction aperture in the full band (0.5-8 keV; Broos \e 2011). Column 6:  Gaussian error on the net X-ray counts. Column 7: A set of mutually exclusive classification hypotheses defined for each source in Broos \e (2011)Ñ H0 : unclassified; H1: source is a foreground main-sequence star; H2: source is a young star, assumed to be in the Carina complex; H3: source is a Galactic background main-sequence star; H4: source is an extragalactic source. Column 8: The sub-cluster within Trumpler~16.  C3, C4, C6 etc (Paper~I).  'Matrix' means no sub-cluster and not in the Southeastern extension, SEM means no sub-cluster and in the Southeastern extension.}
\end{deluxetable}

\begin{deluxetable}{lrcccccccc}
\tabletypesize{\scriptsize}
\tablecaption{XPHOT Derived Properties of Chandra ACIS Point Sources in the Trumpler~16  Region \label{xphot.tab}}
\tablewidth{0pt}
\tablehead{
\colhead{Seq. \#} &\colhead{Designation}&\colhead{Median} &\colhead{error} &
\colhead{Log flux}&  \colhead{Statistical err} &\colhead{Systematic err} &
\colhead{Log \nh}&  \colhead{Statistical err} &\colhead{Systematic err} \\
\colhead{~} &\colhead{~ }&\colhead{energy [keV]} &\colhead{Med. energy} &
\colhead{[ergs cm$^2$ sec$^{-1}$]}&  \colhead{log flux} &\colhead{log flux} &
\colhead{[cm$^{-2}$]}&  \colhead{Log \nh} &\colhead{Log \nh} \\
 }
\startdata
5277 &  104405.29$-$594543.4   &  0.97  &   0.12  & $-$14.64  & $-$15.19  & $-$16.33  &  20.26   &  0.00  &   0.26\\ 
5294 &  104405.79$-$594353.6   &  1.08  &   0.19  & $-$14.87  & $-$15.21  & $-$16.56  &  20.26   &  0.36  &   0.26\\ 
5409 &  104407.86$-$594315.7   &  2.84  &   0.24  & $-$13.23  & $-$13.98  & $-$13.97  &  22.41   &  0.09  &   0.05\\ 
5416 &  104408.06$-$594522.4   &  1.76  &   0.16  & $-$13.95  & $-$14.56  & $-$14.50  &  21.95   &  0.15  &   0.09\\ 
5466 &  104409.04$-$594538.7   &  1.61  &   0.14  & $-$13.90  & $-$14.55  & $-$14.46  &  21.78   &  0.18  &   0.12\\ 
5493 &  104409.75$-$594338.7   &  1.26  &   0.15  & $-$14.35  & $-$14.78  & $-$14.72  &  20.98   &  0.61  &   0.50\\ 
5496 &  104409.80$-$594448.0   &  1.48  &   0.13  & $-$13.97  & $-$14.60  & $-$14.68  &  21.60   &  0.24  &   0.12\\ 
5534 &  104410.39$-$594311.1   &  1.51  &   0.03  & $-$11.24  & $-$12.63  & $-$12.21  &  21.48   &  0.06  &   0.12\\ 
5541 &  104410.47$-$594352.7   &  1.88  &   0.28  & $-$14.17  & $-$14.56  & $-$14.71  &  22.08   &  0.21  &   0.08\\ 
5576 &  104411.23$-$594445.7   &  1.35  &   0.24  & $-$14.65  & $-$14.91  & $-$15.00  &  21.48   &  0.85  &   0.22\\ 
5607 &  104411.88$-$594223.2   &  1.47  &   0.28  & $-$14.60  & $-$14.87  & $-$14.89  &  21.60   &  0.68  &   0.24\\ 
5609 &  104411.89$-$594414.8   &  1.93  &   0.44  & $-$14.23  & $-$14.54  & $-$14.48  &  22.15   &  0.31  &   0.11\\ 
5629 &  104412.43$-$594212.6   &  1.99  &   0.31  & $-$14.10  & $-$14.54  & $-$14.58  &  22.15   &  0.21  &   0.08\\ 
5638 &  104412.53$-$594351.3   &  1.29  &   0.16  & $-$14.46  & $-$14.90  & $-$14.85  &  21.30   &  0.67  &   0.35\\ 
5647 &  104412.84$-$594333.1   &  1.42  &   0.18  & $-$14.20  & $-$14.58  & $-$14.60  &  21.60   &  0.43  &   0.18\\ 
5649 &  104412.86$-$594344.6   &  2.02  &   0.12  & $-$13.18  & $-$14.11  & $-$13.80  &  22.11   &  0.07  &   0.09\\ 

 \enddata
\tablenotetext{a}{Table 2 with complete notes is published in its entirety in the electronic edition of the Astrophysical Journal. A portion is shown here for guidance regarding its form and content.}
\end{deluxetable}

 \begin{deluxetable}{lrrrrr}
\tabletypesize{\normalsize}
\tablecaption{Summary of Metric for Each Sub-cluster \label{summetric}}
\tablewidth{0pt}
\tablehead{
\colhead{Sub-cluster} &\colhead{No. Sources}&\colhead{Density} &\colhead{Disk Fraction} &
\colhead{$<$ {\it \~A}$_V >$}&  \colhead{ Abs. $K_S$} 
  \\
\colhead{~} &\colhead{~}   & \colhead{~}   &\colhead{~} & \colhead{25/50/75}   &\colhead{25/50/75}\\
\colhead{~} &\colhead{~}   & \colhead{[src pc$^{-2}$]}   &\colhead{~} & \colhead{percentiles}   &\colhead{percentiles}
 }
\startdata
 All & 1187 & 27 & 8.9$\pm0.9$\%                  & 2.9/3.7/4.8 & 1.25/2.0/2.5\\
 no sub & 506 & \nodata & 6.4$\pm1.2$\%     & 2.8/3.8/5.2 & 1.0/2.0/2.5\\ 
 all N. sub & 525 & \nodata & 8.4$\pm1.4$\% & 2.9/3.6/4.3 & 1.25/2.0/2.5\\ 
 3 & 33 & 33 & 14.8$\pm7.4$\%                      & 3.0/3.8/4.5 & 1.75/2.0/2.25\\
 4 & 11 & 34 & 25.0$\pm17.7$\%                    & 2.7/3.5/3.8 & 1.5/2.5/2.5\\
 6 & 109 & 45 & 6.8$\pm2.8$\%                     & 2.9/3.6/4.1 & 1.25/2.25/2.75\\
 9 & 53 & 48 & 4.4$\pm3.1$\%                       & 2.8/3.7/4.3 & 1.75/2.25/3.25\\
 10 & 82 & 42 & 5.4$\pm2.7$\%                     & 3.1/3.7/4.5 & 1.0/2.0/2.5\\
 11 & 71 & 27 & 7.0$\pm3.5$\%                     & 1.9/2.9/3.9 & 1.25/1.75/2.5\\   
 12 & 166 & 42 & 10.6$\pm2.7$\%                & 3.4/3.8/4.3 & 1.25/2.0/2.5\\  
 SE ext & 116 & \nodata & 17.8$\pm4.2$\% & 3.2/4.8/6.3 & 0.5/1.5/2.5\\ 
 14 & 40 & 10 & 21.2$\pm8.0$\%                  & 5.1/7.4/10.5 & 0.25/0.75/1.5\\  
 SE all & 156 & 9 & 18.7$\pm3.7$\%            & 4.3/5.6/8.6 & 0.5/1.25/2.25\\  
 \enddata
\end{deluxetable}

\begin{deluxetable} {lrrrllrr}
\tabletypesize{\scriptsize}
\tablecaption{CCCP Detected OB Stars in the Trumpler 16 Region $^1$ \label{HMSX}}
\tablewidth{0pt}
\tablehead{
\colhead{Seq. \#} &  \colhead {sub Cl.}&\colhead{R.A.} &\colhead{Dec.} &
\colhead{Name}    &  \colhead{SpType} &\colhead{V} &\colhead{X-ray Cts}\\
\colhead{~} &          \colhead {~}&\colhead{(J2000.)} &\colhead{(J2000.)} &
\colhead{~}& \colhead{~}&\colhead{(mag)} &\colhead{(0.5-8.0 keV)}
 }
\startdata
5294 & Matrix & 10 44 05.82 & $-$59 35 11.7 & Cl* Trumpler 16 MJ 224 & B1V & 11.14 & 12.4\\
5534 & Matrix & 10 44 10.39 & $-$59 43 11.1 & WR 25 & WN6h + OB? & 8.1 & 24609\\
5665 & C3 & 10 44 13.20 & $-$59 43 10.2 & Cl* Trumpler 16 MJ 257 & O3/4If & 10.8 & 359.2\\
6676 & C6 & 10 44 32.34 & $-$59 44 31.0 & HD 93204 & O5.5V((f)) & 8.42 & 310.8\\
6773 & C6 & 10 44 33.74 & $-$59 44 15.5 & HD 93205 & O3.5V((f+)) + O8V & 7.75 & 1408.7\\
6955 & Matrix & 10 44 36.70 & $-$59 47 29.7 & Cl* Trumpler 16 MJ 359 & O8V & 10.89 & 68.9\\
6691 & Matrix & 10 44 37.17 & $-$59 40 01.3 & Cl* Trumpler 16 MJ 357 & B0.5V & 11.57 & 6.1\\
7224 & Matrix & 10 44 40.99 & $-$59 40 10.2 & Cl* Trumpler 16 MJ 372 & B0V & 11.4 & 14.2\\
7277 & Matrix & 10 44 41.80 & $-$59 46 56.4 & Cl Trumpler 16 100 & O5.5V & 8.6 & 976\\
7621 & Matrix & 10 44 47.31 & $-$59 43 53.2 & CD$-$59 3303 & O7V + O9.5V + B0.2IV & 8.8 & 108.4\\
8036 & Matrix & 10 44 54.06 & $-$59 41 29.4 & Cl* Trumpler 16 MJ 427 & B1V & 10.9 & 173.1\\
8380 & C12 & 10 44 59.90 & $-$59 43 14.8 & Cl Trumpler 16 26 & B1.5V & 11.66 & 9.5\\
8579 & C11 & 10 45 03.16 & $-$59 40 12.5 & Cl* Trumpler 16 MJ 467 & B0.5V & 10.82 & 6.6\\
 \nodata&\nodata  & 10 45 03.55 & $-$59 41 04.0 & $\eta$ Carin\ae\  & pec. & 6 & \\
8648 & C11 & 10 45 04.78 & $-$59 40 53.5 & Cl Trumpler 16 64 & B1.5V:b & 10.7 & 281.2\\
8705 & C10 & 10 45 05.80 & $-$59 45 19.6 & Cl* Trumpler 16 MJ 484 & O7V & 10 & 204.9\\
8707 & C12 & 10 45 05.83 & $-$59 43 07.7 & Cl* Trumpler 16 MJ 481 & O9.5V & 9.77 & 102.2\\
8714 & C11 & 10 45 05.92 & $-$59 40 05.9 & HD 303308 & O4V((f)) & 8.17 & 1654\\
8758 & C12 & 10 45 06.72 & $-$59 41 56.6 & Cl* Trumpler 16 MJ 488 & O8.5V & 9.9 & 91.8\\
8831 & C11 & 10 45 08.23 & $-$59 40 49.4 & CPD$-$59 2628 & O9.5V + B0.3V & 9.5 & 65.5\\
8832 & C10 & 10 45 08.24 & $-$59 46 07.0 & Cl* Trumpler 16 MJ 496 & O8.5V & 10.93 & 1909.4\\
9028 & C10 & 10 45 12.22 & $-$59 45 00.4 & HD 93343 & O7V(n) & 9.7 & 204.9\\
9038 & C12 & 10 45 12.65 & $-$59 42 48.7 & Cl* Trumpler 16 MJ 513 & B2:V & 11.2 & 14.9\\
9044 & C10 & 10 45 12.72 & $-$59 44 46.2 & CPD$-$59 2635 & O8.5 & 9.3 & 384.1\\
9050 & Matrix & 10 45 12.87 & $-$59 44 19.3 & CPD$-$59 2636 & O8.5 & 9.3 & 579.7\\
9195 & C12 & 10 45 16.52 & $-$59 43 37.0 & Cl Trumpler 16 112 & O4.5((f)) & 9.3 & 624.4\\
9344 & C12 & 10 45 20.57 & $-$59 42 51.2 & Cl* Trumpler 16 MJ 554 & O8.5V & 10.09 & 77.1\\
9857 & C14 & 10 45 36.32 & $-$59 48 23.2 & Cl* Trumpler 16 MJ 596 & O5.5Vz + O9.5V & 12.1 & 69\\
10748 & Matrix & 10 46 05.70 & $-$59 50 49.4 & LS 1886 & O4V & 10.7 & 544.6\\
\enddata
\tablenotetext{1}{Adapted from Skiff (2010)}
\end{deluxetable}

\begin{deluxetable} {llrrrrrrcc}
\tabletypesize{\scriptsize}
\tablecaption{IR colors of CCCP Detected OB Stars in the Trumpler 16 Region$^1$ \label{HMSIR}}
\tablewidth{0pt}
\tablehead{
\colhead{Seq. \#} &  
\colhead{Name}    & 
\colhead{J} &\colhead{Jerr}   & \colhead{H}   &\colhead{Herr} & \colhead{K}   &\colhead{Kerr} &
\colhead{$A_V (J-K)$} &\colhead{$A_V (J-H)$} 
 }
\startdata
5294 & Cl* Trumpler 16 MJ 224 & 15.459 & 0.002 & 14.398 & 0.001 & 13.712 & 0.001 & 8.88 & 8.69 \\
5534 & WR 25 & 6.26 & 0.007 & 5.97 & 0.023 & 5.721 & 0.015 & 3.37 & 3.16 \\
5665 & Cl* Trumpler 16 MJ 257 & 7.84 & 0.013 & 7.381 & 0.037 & 7.061 & 0.017 & 4.46 & 4.37 \\
6676 & HD 93204 & 8.026 & 0.021 & 7.987 & 0.035 & 7.97 & 0.029 & 1.17 & 1.36\\ 
6773 & HD 93205 & 7.389 & 0.009 & 7.386 & 0.027 & 7.342 & 0.029 & 1.13 & 1.10 \\
6955 & Cl* Trumpler 16 MJ 359 & 9.384 & 0.017 & 9.142 & 0.019 & 9.007 & 0.018 & 2.63 & 2.81\\ 
6691 & Cl* Trumpler 16 MJ 357 & \nodata & \nodata & \nodata & \nodata & \nodata & \nodata & \nodata & \nodata\\
7224 & Cl* Trumpler 16 MJ 372 & 10.106 & 0.021 & 9.898 & 0.019 & 9.866 & 0.02 & 2.01 & 2.57 \\
7277 & Cl Trumpler 16 100 & 7.798 & 0.015 & 7.735 & 0.035 & 7.639 & 0.025 & 1.64 & 1.53 \\
7621 & CD$-$59 3303 & 8.343 & 0.013 & 8.344 & 0.021 & 8.286 & 0.019 & 1.17 & 1.07 \\
8036 & Cl* Trumpler 16 MJ 427 & 10.209 & 0.047 & 10.153 & 0.069 & 10.16 & 0.039 & 1.14 & 1.48\\ 
8380 & Cl Trumpler 16 26 & 10.863 & 0.017 & 10.66 & 0.025 & 10.509 & 0.021 & 2.53 & 2.53 \\
8579 & Cl* Trumpler 16 MJ 467 & \nodata & \nodata & \nodata & \nodata & \nodata & \nodata & \nodata & \nodata\\
\nodata & $\eta$ Carin\ae\ & \nodata & \nodata & \nodata & \nodata & \nodata & \nodata & \nodata & \nodata\\
8648 & Cl Trumpler 16 64 & \nodata & \nodata & \nodata & \nodata & \nodata & \nodata & \nodata & \nodata\\
8705 & Cl* Trumpler 16 MJ 484 & 8.649 & 0.019 & 8.416 & 0.017 & 8.335 & 0.021 & 2.34 & 2.75\\
8707 & Cl* Trumpler 16 MJ 481 & 8.916 & 0.007 & 8.8 & 0.021 & 8.762 & 0.02 & 1.61 & 1.91\\
8714 & HD 303308 & 7.707 & 0.013 & 7.714 & 0.033 & 7.625 & 0.045 & 1.29 & 1.03\\
8758 & Cl* Trumpler 16 MJ 488 & 9.45 & 0.015 & 9.412 & 0.031 & 9.36 & 0.027 & 1.32 & 1.35\\
8831 & CPD$-$59 2628 & 9.249 & 0.019 & 9.136 & 0.017 & 9.253 & 0.061 & 0.89 & 1.89\\
8832 & Cl* Trumpler 16 MJ 496 & 9.27 & 0.017 & 8.96 & 0.021 & 8.813 & 0.019 & 3.00 & 3.30\\
9028 & HD 93343 & 8.681 & 0.009 & 8.543 & 0.019 & 8.434 & 0.019 & 2.04 & 2.07\\
9038 & Cl* Trumpler 16 MJ 513 & 9.83 & 0.019 & 9.646 & 0.025 & 9.531 & 0.023 & 2.28 & 2.40\\
9044 & CPD$-$59 2635 & 8.33 & 0.011 & 8.18 & 0.033 & 8.091 & 0.017 & 2.00 & 2.15\\
9050 & CPD$-$59 2636 & 8.076 & 0.017 & 7.894 & 0.031 & 7.756 & 0.017 & 2.37 & 2.38\\
9195 & Cl Trumpler 16 112 & 8.147 & 0.005 & 7.993 & 0.021 & 7.884 & 0.033 & 2.11 & 2.18\\
9344 & Cl* Trumpler 16 MJ 554 & 9.387 & 0.017 & 9.309 & 0.021 & 9.257 & 0.018 & 1.50 & 1.64\\
9857 & Cl* Trumpler 16 MJ 596 & 9.293 & 0.021 & 8.809 & 0.027 & 8.505 & 0.023 & 4.51 & 4.55\\
10748 & LS 1886 & 8.176 & 0.013 & 7.962 & 0.035 & 7.768 & 0.025 & 2.77 & 2.61\\
\enddata
\tablenotetext{1}{Adapted from Skiff (2010)}
\end{deluxetable}

\begin{deluxetable} {lllrrl}
\tabletypesize{\normalsize}
\tablecaption{ High Mass Stars in the Trumpler 16 Region - Not Detected in X-rays$^1$ \label{HMSN}}
\tablewidth{0pt}
\tablehead{
\colhead{R.A.} &\colhead{Dec.} &
\colhead{Name}    &  \colhead{SpType} &\colhead{V}  & \colhead {sub Cl.}\\
\colhead{(J2000.)} &\colhead{(J2000.)} &
\colhead{~}& \colhead{~}&\colhead{(mag)} & \colhead {~}
 }
\startdata
 10 44 13.80  & $-$59 42 57.0  & Cl* Trumpler 16 MJ 261  & B0V &12.1    & Matrix\\
 10 44 14.75  & $-$59 42 51.7  & Cl* Trumpler 16 MJ 263  & B0.5V &11.9   & Matrix\\
 10 44 26.48  & $-$59 41 02.7  & Cl* Trumpler 16 MJ 306  & B1.5V &9.88    & Matrix\\
 10 44 28.98  & $-$59 42 34.2  & Cl* Trumpler 16 MJ 323  & B2V&12.1    & Matrix\\
 10 44 30.49  & $-$59 41 40.5  & Cl* Trumpler 16 MJ 329  & B1V&10.88   & C4\\
 10 44 32.89  & $-$59 40 25.9  & Cl* Trumpler 16 MJ 339  & B1V&10.8   & Matrix\\
 10 44 38.66  & $-$59 48 14.2  & Cl Trumpler 16 20           & B1:V&10.2    & Matrix\\
 10 44 40.32  & $-$59 41 48.8  & Cl* Trumpler 16 MJ 370  & B1V&10.77    & Matrix\\
 10 44 58.79  & $-$59 49 21.1  & Hen 3--480                      & em   &11.5 & Matrix\\
 10 45 00.24  & $-$59 43 34.4  & Cl Trumpler 16 25           & B2V&11.88   & C12\\
 10 45 02.19  & $-$59 42 01.1  & Cl* Trumpler 16 MJ 466  & B1V& 10.96    & C12\\
 10 45 05.18  & $-$59 41 42.4  & Cl* Trumpler 16 MJ 477  & B1V&12.1   & C12/11\\
 10 45 05.88  & $-$59 44 18.9  & Cl* Trumpler 16 MJ 483  & B2V&11.5    & C12\\
 10 45 09.65  & $-$59 40 08.5  & Cl* Trumpler 16 MJ 499  & B2V&12.2   & C11\\
 10 45 09.74  & $-$59 42 57.1  & Cl* Trumpler 16 MJ 501  & B1V&11.69    & C12\\
 10 45 11.18  & $-$59 41 11.2  & Cl* Trumpler 16 MJ 506  & B1V&10.78    & C11\\
 10 45 19.42  & $-$59 39 37.3  & Cl* Trumpler 16 MJ 547  & B1.5V&12.2   & Matrix\\
 10 45 31.86  & $-$59 51 09.4  & BM VII 10                        & S        &13.6    & SEM\\
 10 45 44.61  & $-$59 50 41.1  & FO 16                              & OB-     &12.34 & SEM\\
 10 45 54.80  & $-$59 48 15.4  & Trumpler 16 MJ 633        & em       &13.1   & SEM\\
 10 46 32.62  & $-$59 49 50.0  & HD 93538                       & A5/8    & 9.6     & SEM\\
\enddata
\tablenotetext{1}{Adapted from Skiff (2010)}
\end{deluxetable}

\begin{deluxetable} {l|rrr}
\tabletypesize{\normalsize}
\tablecaption{Comparison of the Three Massive Clusters in the Carina Nebula \label{comp}}
\tablewidth{0pt}
\tablehead{
\colhead{} &\colhead{Trumpler~14$^{1}$} &\colhead{Trumpler~15$^{2}$} & \colhead{Trumpler~16}\\
 }
\startdata
X-Ray Sources:            &            &             &   \\       
~~~Probable Members$^{3}$ :  & 1378                & 829                 & 1187      \\     
~~~HAWK-I Detections$^{4}$ : & 1219                & 748                 & 1050        \\   
 XLF ($\Gamma$)             &  \nodata            &  $\sim-$1.27           &  $\sim-$1.27\\            
Estimated \# Stars ($\pm10\%$) & 12,500$^{5}$     & 5,900               & 14,000   \\
High Mass Stars$^{6}$ &              46    & 24 & 57     \\
 Area     [ pc$^2$]                 &            &              &    \\        
~~~Core:               &  1.4         &  1.4       & \nodata      \\      
~~~ Total:              &   95          &   50          &  80        \\    
 Density  [X-ray sources pc$^{-2}$]                  &             &             &      \\
~~~ Core:               &300   &200   &  \nodata     \\       
~~~ Total:              &14.5  &16.6   &14.8 \\ 
 Disk Fraction [\%]             & 9.7$^{4}$           & 3.8                 & 7.4 --  north \\
                                &                            &                              &17.8 -- SE ext.  \\               
 Age                        & 2-3 Myr$^{4}$        &  5-10 Myr           & 3-4 Myr$^{4}$  \\ 
\enddata
\tablenotetext{1}{Reference: Acsenso \e (2008) unless otherwise noted.}
\tablenotetext{2}{Reference: Wang \e (2011) unless otherwise noted.}
\tablenotetext{3}{Reference: Broos \e (2011)}
\tablenotetext{4}{Reference: Preibisch \e (2011)}
\tablenotetext{5}{See text for this calculation.}
\tablenotetext{6}{Reference: Skiff \e (2010)}
\end{deluxetable}

\begin{thebibliography}{}
\bibitem[Albacete Colombo, M{\'e}ndez, \& Morrell(2003)]{Albacete Colombo et al.(2003)} Albacete Colombo, J.~F., M{\'e}ndez, M., \& Morrell, N.~I.\ 2003, \mnras, 346, 704 

\bibitem[Albacete-Colombo et al.\ (2008)]{Albacete-Colombo et al.(2008)} Albacete-Colombo, J.~F., Damiani, F., Micela, G., Sciortino, S., \& Harnden, F.~R., Jr.\ 2008, \aap, 490, 1055 (AC08)


\bibitem[Alexander et al.(2003)]{Alexander et al.(2003)} Alexander, D.~M., 
et al.\ 2003, \aj, 126, 539 


\bibitem[Ascenso et al.]{Ascenso et al.(2007)} Ascenso, J., Alves, J., Vicente, S., Lago, M.T.V.T.\ 2007  \aap, 476, 199


\bibitem[Brooks et al.(2005)]{Brooks et al.(2005)} Brooks, K.~J., Garay, 
G., Nielbock, M., Smith, N., \& Cox, P.\ 2005, \apj, 634, 436 

\bibitem[Broos et al.(2011a)]{2007ApJS..168..100W}Broos P.~S., \e 2011a CCCP Catalog paper 

\bibitem[Broos et al.(2011b)]{2007ApJS..168..100W}Broos P.~S., \e 2011b CCCP Paper - A Naive Bayes Source ClassiÞer for X-ray Sources

\bibitem[Broos et al.(2010)]{Broos et al.(2010)} Broos, P.~S., Townsley, L.~K., Feigelson, E.~D., Getman, K.~V., Bauer, F.~E., \& Garmire, G.~P.\ 2010, \apj, 714, 1582 

\bibitem[Broos et al.(2007)]{Broos et al.(2007)} Broos, P.~S., Feigelson, E.~D., Townsley, L.~K., Getman, K.~V., Wang, J., Garmire, G.~P., Jiang, Z., \& Tsuboi, Y.\ 2007, \apjs, 169, 353 

\bibitem[Damiani et al.(1997)]{Damiani et al.(1997)} Damiani, F., Maggio, A., Micela, G., \& Sciortino, S.\ 1997, Statistical Challenges in Modern Astronomy II, 417 

\bibitem[Davidson \& Humphreys(1997)]{Davidson et al.(1997)} Davidson, K., \& Humphreys, R.~M.\ 1997, \araa, 35, 1 

\bibitem[DeGioia-Eastwood et al.(2001)]{DeGioia-Eastwood et al.(2001)} DeGioia-Eastwood, K., Throop, H., Walker, G., \& Cudworth, K.~M.\ 2001, \apj, 549, 578 

\bibitem[Evans et al.(2003)]{Evans et al.(2003)} Evans, N.~R., Seward, F.~D., Krauss, M.~I., Isobe, T., Nichols, J., Schlegel, E.~M., \& Wolk, S.~J.\ 2003, \apj, 589, 509 

\bibitem[Feigelson et al.(2005)]{Feigelson et al.(2005)} Feigelson, E.~D, et al.\ 2005, \apjs, 160, 379 


\bibitem[Feigelson et al.(2011)]{Feigelson et al.(2011)} Feigelson E. \e 2011, CCCP Clustering paper (Paper~I)

\bibitem[Feinstein, Marraco, \& Muzzio(1973)]{Feinstein et al.(1973)} Feinstein, A., Marraco, H.~G., \& Muzzio, J.~C.\ 1973, \aaps, 12, 331 

\bibitem[Feinstein(1982)]{Feinstein(1982)} Feinstein, A.\ 1982, \aj, 87, 1012 

\bibitem[Forte(1978)]{1978AJ.....83.1199F} Forte, J.~C.\ 1978, \aj, 83, 1199 

\bibitem[Gagn\'e et al.(2011)]{2011ApJS..168..100W}Gagn\'e, M., \e 2011 CCCP Hot Star paper

\bibitem[Gaviola(1950)]{Gaviola(1950)} Gaviola, E.\ 1950, \apj, 111, 408 

\bibitem[Getman et al.(2005)]{2005ApJS..160..353G} Getman, K.~V., Feigelson, E.~D., Grosso, N., McCaughrean, M.~J., Micela, G., Broos, P., Garmire, G., \& Townsley, L.\ 2005, \apjs, 160, 353 

\bibitem[Getman et al.(2006)]{2006ApJS..163..306G} Getman, K.~V., Feigelson, E.~D., Townsley, L., Broos, P., Garmire, G., \& Tsujimoto, M.\ 2006, \apjs, 163, 306 

\bibitem[Getman et al.(2010)]{Getman et al.(2010)} Getman, K.~V., Feigelson, E.~D., Broos, P.~S., Townsley, L.~K., \& Garmire, G.~P.\ 2010, \apj, 708, 1760 

\bibitem[Haisch 2001]{Haisch 2001}Haisch, K. E., Lada, E.~A., Lada, C.~J.\ 2001 \apj, 553, L153

\bibitem[Herbst(1976)]{1976ApJ...208..923H} Herbst, W.\ 1976, \apj, 208, 923 

\bibitem[Herschel(1847)]{Herschel(1847)} Herschel, J.~F.~W., Sir 1847, {\it Results of Observations Made During the Years 1834, 5, 6, 7, 8 at the Cape of Good Hope} (London: Smith, Elder and Co.)  

\bibitem[Kendall(1938)]{Kendall(1938)} Kendall, M.\ 1938," {\it Biometrika} 30, 81

\bibitem[King (1962)]{King(1962)} King, I. 1962, \aj, 67, 471

\bibitem[Levato \& Malaroda(1982)]{Levato et al.(1982)} Levato, H., \& Malaroda, S.\ 1982, \pasp, 94, 807 

\bibitem[Morrell, Garcia,  \& Levato(1988)]{Morrell et al.(1988)} Morrell, N., Garcia, B., \& Levato, H.\ 1988, \pasp, 100, 1431 

\bibitem[Naz\'e et al. (2011)]{2017ApJS..168..100W}Naz\'e, Y., \e 2011 CCCP O star  paper 

\bibitem[Povich et al. (2011)]{2027ApJS..168..100W}Povich, M. S. \e 2011 CCCP New OB star  paper 

\bibitem[Preibisch \& Feigelson(2005)]{Preibisch et al.(2005)} Preibisch, T., \& Feigelson, E.~D.\ 2005, \apjs, 160, 390 

\bibitem[Preibisch et al.(2005)]{Preibisch et al.(2005)} Preibisch, T., et al.\ 2005, \apjs, 160, 401 

\bibitem[Preibisch et al.(2011)]{2037ApJS..168..100W}Preibisch, T., \e 2011, CCCP HAWK-I paper

\bibitem[Sanchawala et al.(2007)]{2007ApJ...656..462S} Sanchawala, K., Chen, W.-P., Lee, H.-T., Chu, Y.-H., Nakajima, Y., Tamura, M., Baba, D., \& Sato, S.\ 2007, \apj, 656, 462 

\bibitem[Sanchawala et al.(2007)]{2007ApJ...667..963S} Sanchawala, K., et al.\ 2007, \apj, 667, 963 

\bibitem[Siess, Dufour, \& Forestini(2000)]{Siess et al.(2000)} Siess, L., Dufour, E., \& Forestini, M.\ 2000, \aap, 358, 593 

\bibitem[Skiff(2010)]{Skiff(2010)} Skiff, B.~A.\ 2010, VizieR Online Data Catalog, 1, 2023 

\bibitem[Skrutskie et al.(2006)]{2006AJ....131.1163S} Skrutskie, M.~F., et  al.\ 2006, \aj, 131, 1163 

\bibitem[Smith 
\& Brooks(2008)]{2008hsf2.book..138S} Smith, N., \& Brooks, K.~J.\ 2008, Handbook of Star Forming Regions, Volume II, 138 

\bibitem[Smith(2006)]{Smith(2006)} Smith, N.\ 2006, \apj, 644, 1151 

\bibitem[Smith et al.(2003)]{Smith et al.(2003)} Smith, N., Gehrz, R.~D., Hinz, P.~M., Hoffmann, W.~F., Hora, J.~L., Mamajek, E.~E.,  \& Meyer, M.~R.\ 2003, \aj, 125, 1458 

\bibitem[Smith(1987)]{1987MNRAS.227..943S} Smith, R.~G.\ 1987, \mnras, 227, 
943 


\bibitem[Spearman (1904)]{Spearman (1904)} Spearman, C.\ 1904,  {\it Amer. J. Psychol.}, 15, 72

\bibitem[Stelzer et al.(2005)]{Stelzer et al.(2005)} Stelzer, B., Flaccomio, E., Montmerle, T., Micela, G., Sciortino, S., Favata, F., Preibisch, T., \& Feigelson, E.~D.\ 2005, \apjs, 160, 557 

\bibitem[Th\'e et al.(1980)]{1980A&A....89..209T} The, P.~S., Bakker, R., \& Tjin A Djie, H.~R.~E.\ 1980, \aap, 89, 209 

\bibitem[Townsley et al.(2011a)]{2007ApJS..168..100W}Townsley, L., \e 2011a CCCP Introduction paper 

\bibitem[Townsley et al.(2011b)]{2007ApJS..168..100W}Townsley, L., \e 2011b CCCP Diffuse emission paper

\bibitem[Walborn(1971)]{Walborn(1971)} Walborn, N.~R.\ 1971, \apjl, 167, L31 

\bibitem[Walborn(1973)]{Walborn(1973)} Walborn, N.~R.\ 1973, \apj, 179, 517

\bibitem[Wang et al.(2008)]{2008ApJ...675..464W} Wang, J., Townsley, L.~K., Feigelson, E.~D., Broos, P.~S., Getman, K.~V., Rom{\'a}n-Z{\'u}{\~n}iga, C.~G., \& Lada, E.\ 2008, \apj, 675, 464 

\bibitem[Wang et al.(2007)]{2007ApJS..168..100W} Wang, J., Townsley, L.~K., Feigelson, E.~D., Getman, K.~V., Broos, P.~S., Garmire, G.~P., \& Tsujimoto, M.\ 2007, \apjs, 168, 100 

\bibitem[Wolk et al.(2010)]{Wolk et al.(2010)} Wolk, S.~J., Winston, E., Bourke, T.~L., Gutermuth, R., Megeath, S.~T., Spitzbart, B.~D., \& Osten, R.\ 2010, \apj, 715, 671


\end{thebibliography}
\end{document}